\documentclass[11pt]{article}
\usepackage[margin=1in]{geometry}
\usepackage{natbib}
\usepackage{amssymb}
\usepackage{amsmath}
\usepackage{amsthm}
\newtheorem{proposition}{Proposition}
\usepackage[hyphens]{url}
\usepackage{hyperref}
\IfFileExists{xurl.sty}{\usepackage{xurl}}{}
\usepackage[utf8]{inputenc}
\usepackage{graphicx}
\usepackage{longtable}
\usepackage{booktabs}
\usepackage{threeparttable}
\usepackage{tikz}
\usetikzlibrary{arrows.meta, positioning, fit, backgrounds}
\providecommand{\tightlist}{\setlength{\itemsep}{0pt}\setlength{\parskip}{0pt}}
\newenvironment{keywords}{\par\medskip\noindent\textbf{Key words:} }{\par\medskip}
\hypersetup{colorlinks=true, linkcolor=blue, citecolor=blue, urlcolor=blue}
\usepackage{fancyhdr}
\fancypagestyle{plain}{\fancyhf{}\fancyfoot[C]{\thepage}}
\date{}

\title{When Does Trial--Real-World Data Fusion Improve Precision?\\
Model Auditing and Selection-Aware Inference for Adaptive-TMLE}

\author{M.\ Ehsan Karim\thanks{\texttt{ehsan.karim@ubc.ca}. ORCID: \href{https://orcid.org/0000-0002-0346-2871}{0000-0002-0346-2871}}\\
School of Population and Public Health, University of British Columbia,\\
Vancouver, British Columbia, Canada, and\\
Centre for Advancing Health Outcomes, St.~Paul's Hospital,\\
Vancouver, British Columbia, Canada}

\begin{document}
\maketitle

\begin{abstract}
\noindent
Augmenting a randomized controlled trial (RCT) with real-world data (RWD) promises greater efficiency. How
much a given fusion actually delivers, and how to attach honest uncertainty to that gain, are rarely
characterized. We use adaptive targeted maximum likelihood estimation (A-TMLE) as a worked example of an
estimator that learns a working model and then debiases it, and we develop three reproducible tools for reliable
evidence from combined trial and real-world data. The first is a report card that makes the estimator's
data-adaptively learned bias model auditable. On simulated data it measures how well the model recovers the
shape of the true enrollment-effect surface and attributes the estimator's variance to its structural parts. The second is a
map of when fusion helps versus hurts, benchmarked against a matched trial-only estimator. In the main constant-enrollment simulation grid, bias magnitude was the dominant axis of the efficiency gain, while functional complexity had a smaller effect overall but mattered increasingly at larger bias; an exact population-oracle variance identity, derived under a restricted working model, accounts for the magnitude effect, with the remaining channels confirmed empirically. The gain crosses break-even near a moderate bias and erodes as the trial
grows, so the advantage is finite-sample rather than a form of super-efficiency. The third is selection-aware
inference for the gain, treated as a data-adaptive estimand: the naive standard error undercovers, and among
ten candidate standard errors, only the block jackknife achieved consistently near- or above-nominal coverage in the studied simulations, although its intervals were conservative. Across six
fusions of three openly available trials (four using constructed within-trial external arms, two using genuine non-experimental controls) --- a biomedical HIV trial, a public-health trial, and a job-training
trial --- in only one fusion does the lower block-jackknife limit lie slightly above one, and there only marginally. In the remaining five, the resulting
conclusion is that fusion has not earned an efficiency claim over the RCT alone. These public examples do not include the large, strongly biased external cohorts for which A-TMLE's bias-correction strategy may be most consequential. On real data the toolkit therefore functions mainly as a
guardrail. The learned-model dimension is a stress diagnostic, not a proxy for ground truth, and the
block-jackknife interval decides whether fusion or the RCT-only analysis should be primary.
\end{abstract}


\section{Introduction}\label{sec:intro}

\paragraph{Background and motivation.} The 21st Century Cures Act and the U.S.\ Food and Drug Administration's
Real-World Evidence program have made the augmentation of randomized controlled trials (RCTs) with real-world
data (RWD) a central methodological problem \citep{cures2016_act, fda2018_rwe}. The appeal is efficiency: a trial
delivers unbiased treatment effects but is expensive and small, while observational RWD is abundant but
confounded. A large literature now combines the two --- by reweighting for representativeness, by test-then-pool
borrowing, or by shrinkage --- and a recurring promise is that fusion ``buys'' efficiency over using the trial
alone \citep{colnet2024_review}. Two questions, however, are seldom answered for any given method: \emph{how much}
efficiency does the fusion actually deliver in finite samples, and how does one form an \emph{honest confidence
statement} about that gain --- an interval whose coverage accounts for the data-adaptive selection of the working model rather than conditioning on it as if fixed? Adaptive-TMLE (A-TMLE) of \citet{vdlaan2026_atmle} --- a member of the targeted maximum likelihood estimation
framework \citep{vdlaan2006_tmle} --- is an attractive entry into this
space and our running example throughout. It targets the trial-respecting ATE under only the assumptions known to
hold for an RCT, plus a trial-enrollment positivity condition. Rather than assume exchangeability between the
trial and the real-world data, it integrates the two through a \emph{learned bias model}. The ATE is written as a
pooled estimand minus a bias estimand, namely the conditional effect of trial enrollment on the outcome. That
bias estimand is fit by a data-adaptively selected, relaxed highly-adaptive-lasso (HAL) working model. The method
is root-$n$ consistent and asymptotically normal, and its theory establishes an asymptotic, oracle-conditional super-efficiency in
which the efficiency gain is driven by the \emph{complexity}, not the \emph{magnitude}, of the bias, so that
``even if the bias is large in magnitude, provided it is a simple function $\ldots$ A-TMLE is still expected to
yield efficiency gain'' \citep{vdlaan2026_atmle}.

\paragraph{A-TMLE as a representative adaptive-debiasing problem.} Although A-TMLE uses a particular
trial-plus-RWD decomposition, its inferential architecture is shared by a broader class of orthogonalized
estimators. Double/debiased machine learning and adaptive debiased machine learning first estimate flexible
nuisance functions or a data-driven working submodel, then use sample splitting or cross-fitting and an
orthogonal or targeted correction to reduce first-order regularization bias \citep{chernozhukov2018_dml,
vdlaan2023_adml, hubbard2016_dataadaptive}. Sample splitting can protect inference for the target parameter
under appropriate rate and regularity conditions. It does not, by itself, make the learned representation
scientifically interpretable, nor quantify the uncertainty in a performance functional computed after model
selection. The two questions we study are therefore structural, not peculiar to any one working-model class or to A-TMLE. Whenever a
learned representation determines how information is borrowed or corrected, that representation should be
inspectable. Whenever a finite-sample efficiency gain is computed after selecting that representation, the
uncertainty calculation must account for the selection. Our formal and empirical results remain specific to
A-TMLE, but the audit principle and the need for selection-aware performance inference apply to any
adaptive/debiased estimator with these two features.

\paragraph{The two gaps.} Two gaps stand between A-TMLE's theoretical promise and its reliable use.
\begin{itemize}
\setlength{\itemsep}{0.35em}
\item \emph{The learned object is rarely reported.} The learned bias model is the very object the theory says
drives the gain, yet it is not usually reported, plotted, or audited. A reader typically sees only a final
estimate and a confidence interval. The same reporting gap arises in any adaptive/debiased procedure in which a
learned representation governs borrowing or correction but is treated as an inferential nuisance rather than a
reportable artifact \citep{vdlaan2023_adml, hubbard2016_dataadaptive}.
\item \emph{The gain is characterized only qualitatively.} The ``complexity not magnitude'' claim is asymptotic
and oracle-conditional, and the original authors note that its finite-sample size is left open: ``it remains
unclear in practice how much efficiency gain one should expect $\ldots$ [in] some [cases] the efficiency gain is
small'' \citep{vdlaan2026_atmle}. There is, moreover, no standard error for the efficiency gain itself. It is
reported only as a Monte Carlo ratio across simulation runs.
\end{itemize}

\paragraph{Contributions.} Using A-TMLE as a fully worked case study, we provide three tools for evaluating
adaptive RCT$+$RWD fusion. Table~\ref{tab:contributions} summarizes the claims, evidence, and scope boundary,
and Web Appendix~B gives the extended comparison with related work.
\begin{itemize}
\setlength{\itemsep}{0.45em}
\item \textbf{Audit the learned correction.} We introduce a report card that makes the realized bias working
model visible. In simulations it evaluates recovery of the enrollment-effect surface, attributes the
influence-curve variance to the pooled and bias-correction components, and flags drift between the
cross-validation-selected and targeted working models. On real data, where the truth is unknown, it reports the
effective dimension and the variance attribution. Relative to adaptive debiased machine learning
\citep{vdlaan2023_adml}, our contribution is to \emph{audit} the realized working model rather than only define
and debias its oracle projection. The same audit also maps a safe operating envelope for A-TMLE's own interval. Under
selective trial enrollment, its targeting-drift diagnostic flags the one corner where main-terms nuisances leave
the interval undercovering, and a Super-Learner nuisance fit is needed to restore it (Section~\ref{sec:envelope}).
\item \textbf{Map when fusion improves precision.} We characterize the finite-sample gain jointly over bias
magnitude, functional complexity, the RCT-only reference estimator, the amount of external data, and sample
size. An exact population-oracle identity under a restricted working model explains why bias magnitude is the
dominant variance axis, and the simulations quantify the finite-sample contribution of model complexity
(Section~\ref{sec:theory}). Relative to compatibility testing \citep{yang2023_elastic} or global shrinkage, our
contribution is to show \emph{when} and \emph{by how much} borrowing helps or hurts. The main efficiency map is established under constant trial-enrollment probability; a separate robustness grid examines $W$-dependent enrollment.
\item \textbf{Quantify uncertainty after model selection.} We treat the efficiency gain as a data-adaptive
performance estimand. Standard influence-function, cross-fit, CV-TMLE, ridge, winsorized, and HulC constructions
are anti-conservative or biased in the studied design. A delete-a-fold block jackknife that re-selects and
re-targets the working model is the only candidate with near- or above-nominal coverage across the main grid and
the robustness slices \citep{hubbard2016_dataadaptive}. This is an empirical calibration result for A-TMLE, not a
general theorem for all adaptive/debiased estimators.
\end{itemize}

\begin{table}[htbp]
\centering
\small
\caption{\textbf{The three contributions at a glance:} a methodological reporting framework and A-TMLE-specific empirical results in the studied designs.}
\label{tab:contributions}
\begin{tabular}{@{}p{0.5cm} p{3.0cm} p{5.0cm} p{4.6cm} l@{}}
\toprule
 & Contribution & What it delivers & Headline evidence & \S \\
\midrule
1 & Report card for a learned working model & Makes the opaque bias model $\widehat\tau_S$ auditable: recovery, influence-curve variance attribution, drift & $\mathrm{cor}^\star\!\ge\!0.82$ at large bias; $\mathrm{var}(D_A)$ inflates $6$--$7.5\times$ & \ref{sec:sim-reportcard} \\
\addlinespace[2pt]
 & \quad\emph{applied:} a safe operating envelope & The drift diagnostic flags the one corner where A-TMLE's \emph{own} ATE interval undercovers and should defer to RCT-only, and shows a Super-Learner nuisance fit restores it & coverage $0.47\!\to\!0.94$ at the wiggly corner once the nuisances are flexible & \ref{sec:envelope} \\
\addlinespace
2 & Efficiency map & When the fusion's finite-sample variance gain materializes, as a function of bias magnitude $\times$ complexity $\times$ reference $\times$ $n$ & gain $1.15\!\to\!0.34$; crosses $1$ near $m\!\approx\!1$; erodes with $n$ & \ref{sec:sim-map} \\
\addlinespace
3 & Selection-aware inference & Selection-aware CIs for the efficiency gain; across our grid the block jackknife is the only candidate with consistently near- or above-nominal coverage across the main grid & jackknife cover $0.98$--$1.00$ (conservative); cross-fit/CV-TMLE/HulC undercover the fixed truth ($\le 0.84$); even the best rival, the naive SE, reaches only $0.87$ & \ref{sec:sim-selse} \\
\bottomrule
\end{tabular}
\end{table}

\paragraph{Positioning.} This paper sits at the intersection of RCT$+$RWD fusion \citep{colnet2024_review,
yang2023_elastic, yangding2020_multiobs, rosenman2023_shrinkage, viele2014_historical, dang2025_escvtmle}, adaptive/debiased machine learning with
data-adaptive target parameters \citep{vdlaan2023_adml, hubbard2016_dataadaptive, li2025_reghal}, and
selection-aware inference and resampling \citep{berk2013_posi, chernozhukov2018_dml, zheng2011_cvtmle, kuchibhotla2024_hulc,
efron1981_jackknife, quenouille1956_jackknife}. Where the dominant paradigms --- test-then-pool, shrinkage, and
Bayesian dynamic borrowing \citep{ibrahim2000_powerprior, schmidli2014_map, pocock1976_historical} --- borrow
all-or-nothing or through a global weight, A-TMLE decomposes the ATE and borrows direction-by-direction through an
explicitly learned bias function that, like the rest of this lineage, it does not report. The closest
selection-aware fusion method is experiment-selector CV-TMLE \citep{dang2025_escvtmle}. It data-adaptively
selects between the trial-only and pooled estimands on each cross-validation fold. It does not, however, quantify
how uncertain the resulting efficiency gain is, nor audit the learned bias object. These are the two gaps our
contributions target.
Three real-world-evidence themes run through the contributions: \emph{transportability}, because the learned bias
model is the enrollment-effect surface such an analysis would model \citep{stuart2011_generalizability,
westreich2017_transportability, dahabreh2020_extending}; \emph{uncertainty quantification}, because the
selection-aware standard error supplies it for the efficiency gain; and \emph{reliability of AI/ML methods},
because the report card turns an otherwise-opaque learned component into an inspectable one. The extended
positioning relative to prior work is in Web Appendix~B.

\paragraph{Aims and roadmap.} We stress throughout that an efficiency comparison is \emph{reference-dependent}: our
claims are relative to a matched RCT-only estimator. The work is a rigorous, reproducible
characterization of when and how much fusion helps, and of how to report it. It strengthens A-TMLE in practice: the
report card corroborates its inference across the map and flags the one selective-enrollment corner where its
interval needs a Super-Learner nuisance fit to stay calibrated (Section~\ref{sec:envelope}). Section~\ref{sec:setup} fixes notation, the estimand, and the
identification assumptions; Section~\ref{sec:methods} defines the estimator, the report-card diagnostics, and the
ten selection-aware standard errors; Section~\ref{sec:sim} reports the simulation study (the report card, the
efficiency map, the safe operating envelope, and the selection-aware head-to-head); Section~\ref{sec:application} illustrates all three on three
real trials: a biomedical HIV trial, a public-health trial, and a job-training trial. Across the six resulting
fusions the block-jackknife interval keeps the RCT-only analysis primary in five of six; Section~\ref{sec:discussion} discusses
implications and limitations. All results are reproducible from a public harness (Web Appendix~F). \emph{Takeaway:} the three tools --- a report card, an efficiency map, and a
delete-a-fold block jackknife singled out from ten candidate standard errors --- together give a finite-sample
account of when fusion helps, by how much, and with what reliability, strengthening A-TMLE's use in
practice.

\section{Notation, estimand, and assumptions}\label{sec:setup}

\paragraph{Observed data and estimand.}
We observe $O=(S,W,A,Y)\sim P_0$ on $n$ units, where $S\in\{0,1\}$ indicates RCT enrollment ($S=1$ trial,
$S=0$ external RWD), $W$ are baseline covariates, $A\in\{0,1\}$ is treatment, and $Y$ is the outcome. Our target
is the trial-population ATE
\begin{equation}
\psi_0 \;=\; E\big[\,Y(1)-Y(0)\,\big|\,S=1\,\big] \;=\; E_{W\mid S=1}\big[\,E(Y\mid S{=}1,W,A{=}1)-E(Y\mid S{=}1,W,A{=}0)\,\big],
\label{eq:estimand}
\end{equation}
the parameter the implementation computes as ``average over $S=1$'' via the empirical reweighting $S/\!\operatorname{mean}(S)$.
\citet{vdlaan2026_atmle} also study the covariate-pooled variant that averages the conditional effect over the
pooled distribution of $W$. In our \emph{main} data-generating processes the within-trial effect is homogeneous (a constant,
Section~\ref{sec:sim}) and $W$ has the same marginal law in both arms. The two estimands therefore coincide and
equal a constant, so the efficiency comparison is a pure variance contrast rather than a statement about
heterogeneous effects; a heterogeneous-effect relaxation (Web Appendix~E) checks robustness to this. Identification of \eqref{eq:estimand} from the within-trial conditionals requires only consistency (no
interference; $Y=Y(A)$) together with RCT randomization and treatment positivity within the trial. \emph{Fusing} the
external data adds one further condition, which A-TMLE names as its price: \textbf{trial-enrollment positivity}
$0<\Pi(W)\equiv P(S{=}1\mid W)<1$ $P_W$-almost everywhere \citep{vdlaan2026_atmle}. This is a two-sided overlap
condition. Its lower bound $\Pi(W)>0$ requires every covariate value with positive $P_W$-mass to have positive trial-enrollment probability, placing the external covariate support inside the trial's. Its upper bound
($P(S{=}0\mid W)>0$) ensures the external conditional mean entering $\tau_S$ remains identified. In the two-arm case, identifying the
treated bias term $\tau_S(W,1)$ additionally requires arm-specific external overlap $P(A{=}1\mid S{=}0,W)\in(0,1)$
(the two-arm condition stated with \eqref{eq:bias} below). We write $\Pi(W,A)=P(S{=}1\mid
W,A)$ for the (treatment-aware) trial-membership mechanism, $g(W)=P(A{=}1\mid W)$ for the treatment propensity,
and reserve $\Pi(W)$ for the covariate-only enrollment probability of the positivity assumption.

\paragraph{The bias decomposition.}
A-TMLE writes
\begin{equation}
\psi_0 \;=\; \widetilde\Psi(P_0) \;-\; \Psi^{\#}(P_0),
\label{eq:decomp}
\end{equation}
where $\widetilde\Psi$ is the \emph{pooled-ATE estimand} and $\Psi^{\#}$ is a \emph{bias projection} that
subtracts the contribution the (possibly confounded) RWD would inject. The pooled-ATE estimand is the pooled
conditional treatment-effect contrast $E(Y\mid A{=}1,W)-E(Y\mid A{=}0,W)$ averaged over the trial $W$-law, which
A-TMLE estimates via a working model $\tau_A(W)$ for the pooled conditional average treatment effect learned over
RCT$+$RWD (Web Appendix~A). The bias projection is
built from the conditional RCT-enrollment effect on the outcome,
\begin{equation}
\tau_S(W,A) \;=\; E(Y\mid S{=}1,W,A) - E(Y\mid S{=}0,W,A),
\label{eq:taus}
\end{equation}
through the trial-population specialization of the Lemma~2 identity of \citet{vdlaan2026_atmle}
\begin{equation}
\Psi^{\#}(P) \;=\; E_{W\mid S=1}\big[\,\Pi(0\mid W,0)\,\tau_S(W,0) \;-\; \Pi(0\mid W,1)\,\tau_S(W,1)\,\big],
\quad \Pi(0\mid W,A)=P(S{=}0\mid W,A),
\label{eq:bias}
\end{equation}
which reduces to a single term when the external arm contains only controls. When both external arms are present,
the two-term $\Psi^{\#}$ additionally requires external two-arm overlap, $P(A{=}1\mid S{=}0,W)\in(0,1)$ with joint
support, for the bias projection to be well-defined. Crucially, A-TMLE does \emph{not}
assume trial--external mean exchangeability $E(Y_a\mid S{=}1,W)=E(Y_a\mid S{=}0,W)$; $\Psi^{\#}$ absorbs whatever bias --- including
unmeasured confounding --- the pooled projection introduces. This absorbed bias may reflect unmeasured confounding, source or measurement differences between the trial and external data, or working-model error --- unmeasured confounding is only one of its possible sources. The decomposition~\eqref{eq:decomp} is an exact
population identity for the true-contrast $\widetilde\Psi$. The deployed estimator, however, targets the
\emph{projection} of $\widetilde\Psi$ onto its relaxed-HAL working model. It is therefore consistent, and its asymptotic expansion for
$\psi_0$ carries two second-order terms: the working-model approximation error (the gap between this projection and
the true target $\psi_0$) and the standard second-order nuisance-estimation remainder of the underlying TMLE.
Asymptotic linearity requires both to be $o_P(n^{-1/2})$; both are, once the working model approximates the oracle
bias model and the initial nuisance estimators converge, each at rate $n^{-1/4}$, under the empirical-process
(Donsker) and further regularity conditions \citep{vdlaan2026_atmle}.

\begin{figure}[htbp]
\centering
\begin{tikzpicture}[
  font=\small,
  box/.style={draw, rounded corners, align=center, inner sep=4pt, minimum height=8mm},
  >={Stealth[length=2mm]}]
\node[box] (O) {Data $O=(S,W,A,Y)$\\ trial $S{=}1$ $+$ RWD $S{=}0$};
\node[box, above right=3mm and 16mm of O] (A) {Pooled projection $\widetilde\Psi$\\ working model $\tau_A(W)$};
\node[box, below right=3mm and 16mm of O] (Sm) {Bias projection $\Psi^{\#}$\\ \textbf{learned} $\tau_S(W,A)$};
\node[box, right=42mm of O] (psi) {$\widehat\psi=\widetilde\Psi-\Psi^{\#}$\\ EIC $D=D_A-D_S$};
\node[box, right=14mm of psi] (R) {gain $R=\dfrac{\mathrm{var}(D_{\mathrm{rct}})}{\mathrm{var}(D_{\mathrm{atmle}})}$\\ block-jackknife CI};
\draw[->] (O) -- (A); \draw[->] (O) -- (Sm);
\draw[->] (A) -- (psi); \draw[->] (Sm) -- (psi); \draw[->] (psi) -- (R);
\node[box, dashed, below=5mm of Sm, fill=gray!8] (rc) {\textbf{Report card} audits $\tau_S$:\\ recovery $\cdot$ variance attribution $\cdot$ drift};
\draw[->, dashed] (Sm) -- (rc);
\end{tikzpicture}
\caption{\textbf{Schematic.} A-TMLE decomposes the trial-population ATE into a pooled projection $\widetilde\Psi$
(via the working model $\tau_A$) minus a bias projection $\Psi^{\#}$ (via the learned $\tau_S$); the efficient
influence curve is $D=D_A-D_S$. Contribution~(1) audits $\tau_S$ (the report card); contribution~(2) maps the gain
$R$; contribution~(3) builds a selection-aware block-jackknife interval for $R$.}
\label{fig:schematic}

\noindent\emph{Alt text:} Flow diagram. Trial-plus-real-world data $O=(S,W,A,Y)$ feeds two branches --- a pooled
projection $\widetilde\Psi$ using working model $\tau_A$, and a bias projection $\Psi^{\#}$ using a learned
$\tau_S$ --- which combine into the A-TMLE estimate with influence curve $D=D_A-D_S$, from which the efficiency
gain $R$ and its block-jackknife confidence interval are formed; a dashed box shows the report card auditing
$\tau_S$ for recovery, variance attribution, and drift.
\end{figure}
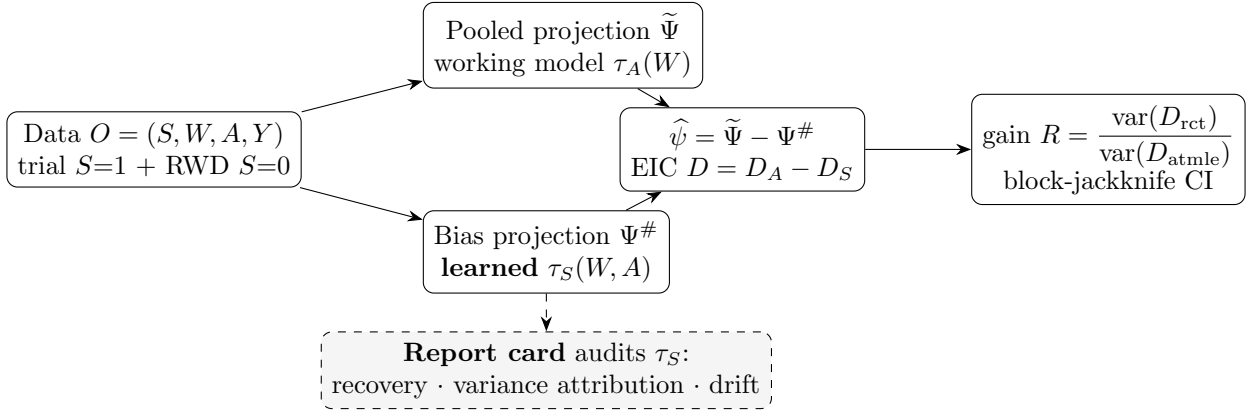

\paragraph{Influence curve and the efficiency gain.}
At the solution the estimator is asymptotically linear with efficient influence curve
\begin{equation}
D \;=\; D_A - D_S,
\label{eq:eic}
\end{equation}
where $D_A$ is the pooled-projection component and $D_S$ the bias-correction component (which carries an extra
trial-membership score reflecting that $\Pi$ is estimated); the reported standard error is
$\widehat{\mathrm{SE}}=\{\widehat{\mathrm{Var}}(D)/n\}^{1/2}$, the empirical influence-curve variance. The object
of Sections~\ref{sec:sim}--\ref{sec:application} is the \emph{efficiency gain}
\begin{equation}
R \;=\; \frac{\operatorname{var}(D_{\mathrm{rct}})}{\operatorname{var}(D_{\mathrm{atmle}})},
\label{eq:gain}
\end{equation}
the ratio of the influence-curve variances of $D_{\mathrm{rct}}$ and of A-TMLE's $D_{\mathrm{atmle}}$, where
$D_{\mathrm{rct}}$ is the influence function of a matched RCT-only estimator, the correctly-specified-GLM
cross-fitted AIPW defined in Section~\ref{sec:sim-design}. Here $R>1$ means A-TMLE is tighter. For a fixed working model, $R$ is a fixed functional of the distribution, but the implemented target is itself a
data-adaptive (and $n$- and reference-dependent) estimand---the working model is cross-validation-selected---and Section~\ref{sec:sim-selse} asks how to put a
calibrated interval on it. Figure~\ref{fig:schematic} summarizes the
decomposition and where each contribution attaches.

\paragraph{Scope caveat.}
The main data-generating processes of Section~\ref{sec:sim} assign trial membership deterministically, so
$\Pi(W)$ is constant and the enrollment-positivity assumption holds trivially; the headline report-card and
efficiency-map conclusions are, strictly, statements about this constant-positivity regime. Selective
($W$-dependent) enrollment is then studied directly rather than left to future work. We relax the efficiency map
to $W$-dependent enrollment (Section~\ref{sec:sim-map}) and map the safe operating envelope for A-TMLE's own
interval (Section~\ref{sec:envelope}), with the full grids in Web Appendix~E.

\section{Methods}\label{sec:methods}

\subsection{The A-TMLE estimator}\label{sec:methods-estimator}
Both working models are estimated by an $R$-learner with a \emph{relaxed-HAL} representation: a HAL basis \citep{benkeser2016_hal, vdlaan2017_haltmle} screened
by an $\ell_1$ penalty with cross-validated penalty, then refit unpenalized on the selected basis. The treatment
main-effect basis is forced into $\tau_S$ unpenalized. The pair $(\Pi,\tau_S)$ together with a targeted
trial-membership mechanism $\Pi^\star$ are updated by iterated TMLE fluctuations, while $\tau_A$ receives a single
fluctuation. The full estimator mechanics --- the $R$-loss, the clever covariates, the fluctuation submodels, and
the reproducibility-critical defaults --- are in Web Appendix~A; the estimator is the \texttt{atmle} package of
\citet{vdlaan2026_atmle}.

\subsection{A report card for the learned working model}\label{sec:methods-reportcard}
We attach three diagnostics to a fitted A-TMLE fusion estimator that make the learned bias object legible.
\begin{itemize}
\setlength{\itemsep}{0.35em}
\item \emph{Recovery} is available only when the data-generating truth is known. It compares the targeted learned
surface $\widehat\tau_S(W,A)$ with $\tau_{S,0}(W,A)$ through $\mathrm{cor}^\star=\mathrm{cor}(\widehat\tau_S,\tau_{S,0})$.
\item \emph{Attribution} decomposes the influence-curve variance $\operatorname{var}(D)$ into the pooled-projection
variance $\operatorname{var}(D_A)$ and the net remainder $\operatorname{var}(D)-\operatorname{var}(D_A)=\operatorname{var}(D_S)-2\operatorname{cov}(D_A,D_S)$ (the bias-correction component together with its covariance with the pooled projection), reported alongside the
gain~\eqref{eq:gain}. It shows whether pooling and bias correction reduce or increase variance relative to the
RCT-only analysis.
\item \emph{Drift} is a $\Pi$-weighted distance between the cross-validation-selected and the TMLE-targeted
$\tau_S$. A large drift indicates that targeting materially changes the selected surface and may signal
instability or model mismatch; it does not by itself identify an over-sparse working model.
\end{itemize}

\paragraph{What the effective dimension means on real data.} When $\tau_{S,0}$ is unknown, recovery is
unavailable, and the report card reports the effective dimension of the targeted relaxed-HAL bias model together
with the variance attribution. The active-basis count is not a recovery score, an estimate of unmeasured
confounding, or a calibrated distance between populations. Because the treatment main-effect basis is forced into
$\tau_S$, a count of one means no additional data-selected basis was retained; larger counts record how many
further directions the data require to represent the residual enrollment effect. A relatively large count
therefore signals that a low-dimensional bridge between the trial and external outcomes was inadequate. It is a
transportability-style diagnostic of how the two populations differ, not a pass/fail test of a required
assumption. A-TMLE does not need trial--external outcome exchangeability to identify the trial ATE. Nor is
the count a stand-alone predictor of efficiency: in the studied finite-sample settings, bias magnitude was the dominant axis of the efficiency gain while functional complexity mattered increasingly only at larger bias
(Section~\ref{sec:theory}), and the effective count can be non-monotone when the true surface is rougher than the
HAL dictionary. We therefore read it as a \emph{stress flag}, interpreted jointly with overlap, variance
attribution, and the selection-aware interval for the gain.

\paragraph{Operational decision rule.}\label{par:decision-rule} For a single fusion analysis we recommend the
following, which bases the efficiency verdict on the gain and its selection-aware interval rather than on the basis count.
\begin{enumerate}
\setlength{\itemsep}{0.3em}
\item Verify treatment and enrollment overlap; if overlap is materially violated, keep the RCT-only analysis as
primary.
\item Report the gain $\widehat R$~\eqref{eq:gain} with its block-jackknife interval in every case.
\item Read a materially larger basis count, relative to a prespecified comparator (for example, a companion external arm believed unbiased, or a prior fusion of comparable sample size), as evidence that borrowing
relies on a more elaborate correction --- a stress flag, not proof that fusion is invalid. When no such comparator is available, as on a first single fusion, omit this step: the remaining steps require none.
\item If $\widehat R\le 1$, or if $\widehat R>1$ but the block-jackknife interval includes one, treat the RCT-only
estimate as the primary result and do not claim an efficiency improvement.
\item Present A-TMLE as an efficiency-improving analysis only when overlap is adequate and the lower block-jackknife limit for $R$ exceeds one, using the variance attribution to explain where the variance change arises rather than as a separate pass/fail criterion. This rule is an empirical recommendation from
the A-TMLE studies here, not a theorem for adaptive/debiased estimators in general.
\end{enumerate}

\subsection{Selection-aware standard errors}\label{sec:methods-selse}
The efficiency gain~\eqref{eq:gain} is itself an estimand, and its naive influence-function SE conditions on the
one cross-validation-selected HAL working model. We benchmark ten SE estimators of $\log(\text{gain})$: the
\textbf{naive} influence-function SE; a delete-a-fold \textbf{block jackknife} that re-selects the working model
on each $S$-stratified leave-fold-out subsample; a \textbf{cross-fitted} out-of-fold influence-curve variance
(with model selection refit within each training fold); three \textbf{ridge} stabilizations of the cross-fit
information-matrix inverse ($\lambda=0.01,0.05,0.2$); a \textbf{winsorized} cross-fit; a principled
\textbf{CV-TMLE} that re-targets a one-dimensional fluctuation for each of $\tau_A$ and $\tau_S$ on each held-out
fold; CV-TMLE plus ridge; and an assumption-lean \textbf{HulC} interval (the convex hull of six disjoint-group
gains, valid under only median-unbiasedness). Each is defined precisely in Web Appendix~C.

\section{Simulation study}\label{sec:sim}
We assess all three contributions on a common simulation grid; Table~\ref{tab:sim_roadmap} is a roadmap of the
study, pointing each component to its headline display in the main text and its full per-cell results in the
Supplementary Material. We pair each of the two headline findings with an analytic account where it arises --- a
population-variance identity for the magnitude-dominance (Section~\ref{sec:theory}) and a heuristic for the
standard-error failures (Section~\ref{sec:sim-selse}).

\begin{table}[htbp]
\centering
\begin{threeparttable}
\caption{\textbf{Roadmap of the simulation study.} Each component, the question it answers and what it shows, and
where to find the headline display in the main text and the full per-cell results in the Supplementary Material
(Web Appendices). Rows follow the reading order of Section~4. The three contributions are (1) the report card,
(2) the efficiency map, and (3) selection-aware inference.}
\label{tab:sim_roadmap}
\small
\begin{tabular}{@{}p{0.21\linewidth} p{0.55\linewidth} p{0.18\linewidth}@{}}
\toprule
Component & Question it answers, and what it shows & Where \\
\midrule
\textbf{(1)} Bias-model report card & \emph{Does the audit recover the learned bias surface?} It recovers the true
enrollment-effect surface and attributes the influence-curve variance to its parts. & Table~\ref{tab:recovery},
Fig.~\ref{fig:recovery}; Web App.~D \\
\addlinespace
\textbf{(2)} Efficiency map & \emph{When does fusion help versus hurt?} The gain is driven by bias \emph{magnitude},
not complexity, crossing parity near a moderate bias. & Table~\ref{tab:gainmap}, Fig.~\ref{fig:gainmap}; Web App.~D \\
\addlinespace
\quad Bias and coverage & \emph{Is the sub-unity gain a variance cost, not bias?} A-TMLE stays approximately
unbiased with near-nominal coverage across the map. & Table~\ref{tab:biascov}; Web App.~D \\
\addlinespace
\quad Reference panel and $n$-ladder & \emph{Is the verdict an artifact of the benchmark, and does it persist as the
trial grows?} It holds against three references; the near-zero-bias gain erodes toward parity as the trial grows, while a biased gain does not. & Table~\ref{tab:refpanel};
Web App.~D \\
\addlinespace
\quad Graded-complexity ladder & \emph{Is ``magnitude, not complexity'' an artifact of the three discrete bias
shapes?} A graded roughness ladder holds magnitude fixed while grading complexity; magnitude stays leading-order. &
Web App.~E \\
\addlinespace
Analytic account (\S\ref{sec:theory}) & \emph{Why does magnitude dominate, and why do the naive standard errors
fail?} The Proposition~\ref{prop:varDA} population identity and a heuristic account, respectively. & \S\ref{sec:theory} and \S\ref{sec:sim-selse}; Web App.~G \\
\addlinespace
Safe operating envelope & \emph{Where is A-TMLE's \emph{own} ATE interval trustworthy?} Nominal across
surfaces and under selective enrollment, except at one non-decaying rough surface at large bias, where the
shortfall deepens with $n$; a Super-Learner nuisance fit restores it. & Table~\ref{tab:envelope},
\S\ref{sec:envelope}; Web App.~E \\
\addlinespace
\textbf{(3)} Selection-aware standard error & \emph{Which standard error for the gain achieves nominal coverage?} Of ten
candidates only the block jackknife attains nominal coverage; naive, cross-fit, CV-TMLE, and HulC all fall short across our grid. &
Table~\ref{tab:selse}, Fig.~\ref{fig:selsecov}; Web App.~D \\
\addlinespace
\quad Robustness slices & \emph{Does the verdict survive selective enrollment, heterogeneity, larger samples, a
binary outcome, and other stresses?} It does, across all slices. & Web App.~E \\
\bottomrule
\end{tabular}
\begin{tablenotes}[flushleft]\footnotesize
\item[] ``Web App.'' refers to the Supplementary Material (Web Appendices A--G), which hold the full per-cell grids,
  the complete ten-method standard-error comparison, all robustness slices, the software/reproducibility map, and
  the proofs.
\end{tablenotes}
\end{threeparttable}
\end{table}

\subsection{Design}\label{sec:sim-design}
We generate RCT$+$RWD data with a true, homogeneous ATE of $1.5$ and a bias that enters only the RWD arm. With
three standard-normal covariates and Gaussian outcome noise, the structural outcome is
$Y = 2.5 + 0.9W_1 + 1.1W_2 + 2.7W_3 + 1.5A + U_Y + \mathbf 1\{S{=}0\}\,B$, so the treatment effect is the constant
$1.5$ in every cell and the bias $B=B(W,A)$ perturbs only the external arm. Treatment is randomized in the trial
($P(A{=}1\mid S{=}1)=0.67$) and confounded externally ($P(A{=}1\mid S{=}0,W)=\operatorname{expit}(0.5W_1)$),
matching the original paper's allocation. The bias is parameterized so that its \emph{magnitude} $m$ (the bias standard deviation, in outcome-residual SD
units) and its \emph{complexity} (functional shape: linear, interaction, or wiggly) vary \emph{orthogonally in the total bias SD}:
every shape is scaled to the same total bias SD $m$, so the two axes are decoupled at the level of that total. The three shapes
do, however, differ in the arm-dependent share that survives the pooled projection, so the graded roughness ladder
(linear $\to$ interaction $\to$ wiggly) is the cleaner evidence on complexity.
Each shape is normalized to unit standard deviation over the external population, then scaled by $m$, i.e.\
$B=m\,s(W,A)/\mathrm{sd}_0(s)$ with $s(W,A)$ the unit-shape carrier and $\mathrm{sd}_0(s)$ its standard deviation
over the external ($S{=}0$) population. Thus $m$ is the bias SD. We
benchmark A-TMLE's efficiency against a \textbf{matched,
correctly-specified-GLM, cross-fitted RCT-only AIPW} --- main-terms logistic propensity and linear outcome
GLMs, cross-fitted, returning an influence curve on the same ``average over $S{=}1$'' scale as A-TMLE (a one-step estimator, asymptotically equivalent to a correctly-specified TMLE with the same GLM nuisances). This is a \emph{method}-matched reference (the same parametric nuisance class A-TMLE is given, the same estimand,
influence-curve-returning), not a learner-vs-learner contrast. Section~\ref{sec:sim-refpanel} shows a genuinely
flexible reference collapses to this GLM on these near-linear DGPs, so it is not a weak denominator. In fact, because
the bias enters only the external arm, on the trial arm ($S{=}1$) the outcome is \emph{exactly} a main-terms linear
model, $\mu(W)=2.5+0.9W_1+1.1W_2+2.7W_3$ with constant treatment effect $1.5$; the matched-GLM AIPW reference is
therefore correctly specified on the trial arm by construction, and, being the \emph{asymptotically} efficient benchmark there, it anchors the
efficiency comparison from the efficient side. Full DGP and
reference definitions are in Web Appendix~C. \emph{Takeaway:} the design isolates efficiency, holding the true
ATE fixed at $1.5$ while sweeping bias magnitude and complexity orthogonally, so any change in the gain is
attributable to the bias rather than to the estimand.

\subsection{The bias-model report card}\label{sec:sim-reportcard}
\begin{figure}[htbp]
\centering
\includegraphics[width=0.78\linewidth]{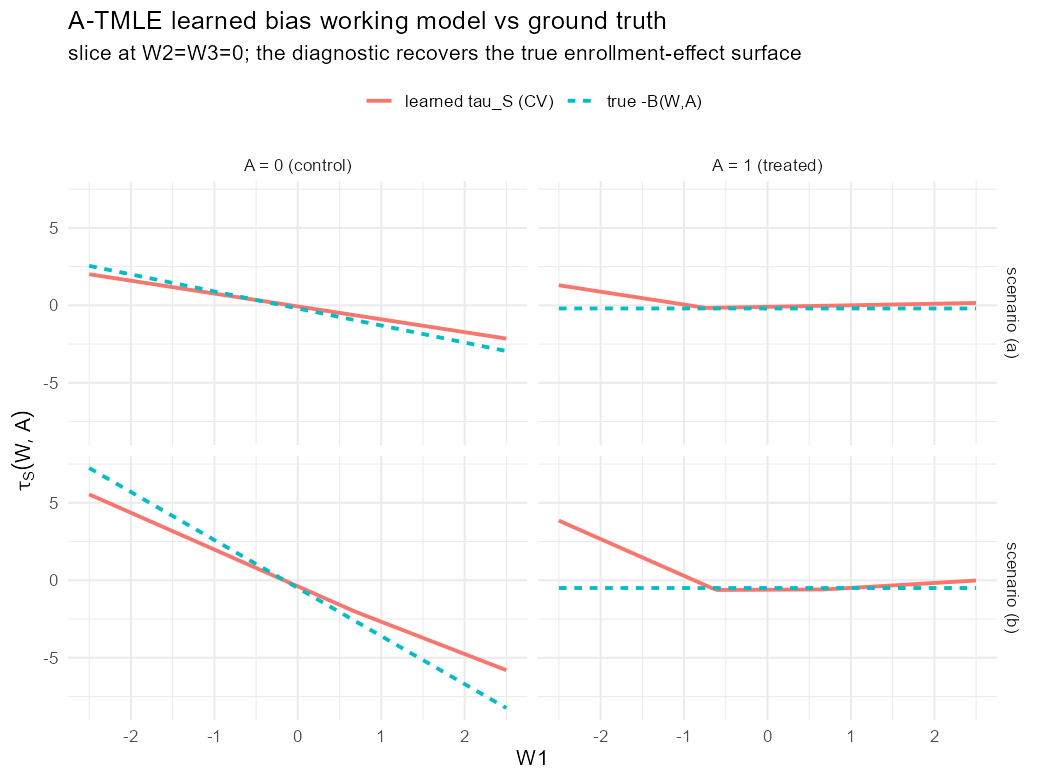}
\caption{\textbf{Recovery surface.} The learned bias model $\widehat\tau_S(W,A)$ (solid, from cross-validated
relaxed-HAL) against the truth $\tau_{S,0}(W,A)=-B(W,A)$ (dashed), sliced at $W_2=W_3=0$, by treatment arm
($A=0$/$A=1$, columns) and scenario (rows). The report card reproduces the structure of the enrollment-effect
surface, including the arm-specific $W_1$-dependence.}
\label{fig:recovery}

\noindent\emph{Alt text:} Grid of line plots. In each panel the learned bias-model surface (solid) closely
tracks the true enrollment-effect surface (dashed) as a function of $W_1$, across treatment arms (columns
$A=0$, $A=1$) and scenarios (rows), sliced at $W_2=W_3=0$, including the arm-specific $W_1$-dependence.
\end{figure}

Table~\ref{tab:recovery} reports recovery and the mean working-model size across the grid ($B=1000$ replicates per
cell; Monte Carlo SE of $\mathrm{cor}^\star\le 0.005$). Recovery is strong where it matters. For every complexity, $\mathrm{cor}^\star\ge 0.82$ once the bias reaches two
residual SDs ($m\ge 2$), and $\approx 0.95$ for a linear bias. It degrades only at small magnitude, where the true
$\tau_S$ is near-constant and the gain is already $\sim 1.15$, so imperfect recovery there is least consequential for the efficiency conclusion. It is undefined
at $m=0$, where there is no bias to recover. On the original paper's own scenarios (a)/(b) the targeted surface attains
$\mathrm{cor}^\star\approx 0.93$/$0.95$ (illustrative, each over $B=8$ diagnostic replicates, Monte Carlo SD $\approx 0.02$), and it
reproduces qualitative structure --- for instance that the $W_1$-dependence of the bias is present in the control
arm and absent in the treated arm (Figure~\ref{fig:recovery}). Because $\mathrm{cor}^\star$ is invariant to affine rescaling, it assesses recovery of the surface's pattern but
not its scale. The small ATE bias in Table~\ref{tab:biascov} provides indirect reassurance about the scale relevant
to $\Psi^{\#}$, but it does not validate the full $\tau_S(W,A)$ surface. \emph{Attribution} shows
$\operatorname{var}(D_A)$ inflates roughly six- to seven-and-a-half-fold across the grid, from $\approx 4.2$ at
$m=0$ to $\approx 25$--$31$ at $m=4$ (by complexity). The dominant cost is therefore pooling a noisier, biased
outcome, not the bias-correction step itself (Web Appendix~D). \emph{Drift} is small in the linear DGPs and
grows monotonically with bias (from $\approx 0.002$ at $m=0$ to $\approx 0.03$--$0.05$ at $m=4$), a quantitative
warning that the working model is being stretched. \emph{Takeaway:} the bias model is recovered well where it matters ($\mathrm{cor}^\star\ge 0.82$ for $m\ge 2$),
making auditable a central data-adaptive object that is not usually reported.

\begin{table}[htbp]
\centering
\begin{threeparttable}
\caption{\textbf{Report card: recovery of the learned bias model $\widehat\tau_S$.} For each bias shape and
magnitude $m$, $\mathrm{cor}^\star$ is the correlation between the targeted learned bias surface and the truth,
and $\bar d$ the mean number of active basis terms in the selected working model, over $B=1000$ replicates.
Recovery is strong where it matters --- $\ge 0.82$ for every shape once $m\ge 2$, $\approx 0.95$ for a linear bias
--- and degrades only at small magnitude, where the true bias is near-constant.}
\label{tab:recovery}
\small
\begin{tabular}{@{}l cc cc cc cc cc@{}}
\toprule
 & \multicolumn{2}{c}{$m=0$} & \multicolumn{2}{c}{$m=0.5$} & \multicolumn{2}{c}{$m=1$} & \multicolumn{2}{c}{$m=2$} & \multicolumn{2}{c}{$m=4$} \\
\cmidrule(lr){2-3}\cmidrule(lr){4-5}\cmidrule(lr){6-7}\cmidrule(lr){8-9}\cmidrule(lr){10-11}
Bias shape & $\mathrm{cor}^\star$ & $\bar d$ & $\mathrm{cor}^\star$ & $\bar d$ & $\mathrm{cor}^\star$ & $\bar d$ & $\mathrm{cor}^\star$ & $\bar d$ & $\mathrm{cor}^\star$ & $\bar d$ \\
\midrule
Linear      & --- & 2.3 & 0.718 & 7.6  & 0.889 & 13.0 & 0.948 & 16.2 & 0.967 & 18.0 \\
Interaction & --- & 2.4 & 0.509 & 8.3  & 0.763 & 17.2 & 0.860 & 26.0 & 0.886 & 31.2 \\
Wiggly      & --- & 2.4 & 0.411 & 7.8  & 0.699 & 19.7 & 0.824 & 30.6 & 0.860 & 38.0 \\
\bottomrule
\end{tabular}
\begin{tablenotes}[flushleft]\footnotesize
\item[] $m$, bias magnitude (the bias standard deviation, in outcome-residual SD units); $m=0$ is unbiased RWD.
\item[] $\mathrm{cor}^\star=\mathrm{cor}(\widehat\tau_S,\tau_{S,0})$, the Pearson correlation of the \emph{targeted}
  learned bias surface $\widehat\tau_S$ with the truth $\tau_{S,0}(W,A)=-B(W,A)$; undefined (---) at $m=0$, where
  there is no bias to recover. Monte Carlo SE of $\mathrm{cor}^\star\le 0.005$.
\item[] $\bar d$, mean number of active highly-adaptive-lasso basis terms in the cross-validation-selected
  working model (an effective dimension of $\widehat\tau_S$).
\end{tablenotes}
\end{threeparttable}
\end{table}

\subsection{The efficiency map}\label{sec:sim-map}
Table~\ref{tab:gainmap} reports the gain~\eqref{eq:gain} over magnitude $\times$ complexity ($B=1000$, $15{,}000$
fits, zero failures). Three findings.

\begin{table}[htbp]
\centering
\begin{threeparttable}
\caption{\textbf{The magnitude $\times$ complexity efficiency map.} Influence-curve efficiency gain $\widehat R$
of A-TMLE relative to a matched, correctly-specified-GLM, cross-fitted RCT-only AIPW, over the bias
magnitude $m$ and bias complexity. $\widehat R>1$ means A-TMLE is tighter than the trial-only estimator. The gain
is driven primarily by magnitude --- falling from $\sim 1.15$ at zero bias to $0.34$--$0.59$ at large bias and
crossing parity just above $m\approx 1$ --- with complexity separating the shapes only at large magnitude. Even a simple
(linear) large bias drives the gain below one.}
\label{tab:gainmap}
\small
\begin{tabular}{@{}l ccccc@{}}
\toprule
Bias shape & $m=0$ & $m=0.5$ & $m=1$ & $m=2$ & $m=4$ \\
\midrule
Linear      & 1.155 & 1.080 & 1.031 & 0.906 & \textbf{0.589} \\
Interaction & 1.150 & 1.100 & 1.012 & 0.757 & 0.386 \\
Wiggly      & 1.150 & 1.114 & 1.018 & 0.732 & 0.340 \\
\bottomrule
\end{tabular}
\begin{tablenotes}[flushleft]\footnotesize
\item[] $\widehat R=\widehat{\mathrm{var}}(D_{\mathrm{rct}})/\widehat{\mathrm{var}}(D_{\mathrm{atmle}})$, the ratio
  of the influence-curve variance of the matched RCT-only estimator to that of A-TMLE (Eq.~\ref{eq:gain}).
\item[] $m$, bias magnitude (bias standard deviation, in outcome-residual SD units). $B=1000$ replicates per cell
  ($15{,}000$ A-TMLE fits, zero failures); $n_{\mathrm{rct}}=n_{\mathrm{ext}}=250$. Monte Carlo SE
  $\approx 0.003$--$0.005$.
\end{tablenotes}
\end{threeparttable}
\end{table}

\begin{enumerate}
\tightlist
\item \textbf{Magnitude is the dominant axis.} The gain falls monotonically with $m$, crossing $1.0$ just above
$m\approx 1$ --- between $m=1$ and $m=2$, a bias of $\sim$one outcome-residual SD --- for every complexity. The
mechanism is transparent. As noted in Section~\ref{sec:sim-reportcard}, the pooled-projection variance
$\operatorname{var}(D_A)$ inflates $\sim 6$--$7.5\times$ across the range; pooling a noisier biased outcome is the
cost, not the bias-correction step.
\item \textbf{Complexity is a second-order amplifier.} At $m\le 1$ the three complexity curves agree to within
$\sim 0.03$; at $m\ge 2$ they fan out (linear $>$ interaction $>$ wiggly), the spread growing from $\sim 0.17$
($m=2$) to $\sim 0.25$ ($m=4$). A graded \emph{roughness ladder} --- varying complexity continuously while holding
magnitude, bias-SD, and the low-order projection fixed (Web Appendix~E) --- confirms this: the gain is flat across
six roughness rungs at $m\le 1$ (spread $\le 0.03$) and fans out only at large bias, where it tracks the
\emph{effective} basis count, which rises with roughness then \emph{saturates and falls} once the oscillation
outruns the HAL knot resolution. Consequently the active-basis count should be read relative to the algorithm and its tuning. It is not a monotone estimator of the true surface complexity, and it does not on its own decide whether fusion helps.
\item \textbf{A simple but large bias does not, on its own, preserve the gain at this sample size.} A linear
bias at $m=4$ gives $0.589$ (MCSE $0.004$), firmly below one. Simplicity of bias shape therefore does not by
itself keep the gain above parity once the bias is large in magnitude, at $n_{\mathrm{rct}}=250$. This reading is
\emph{reference-dependent}: against a matched, efficient RCT-only estimator a simple large bias is a net
efficiency loss. Section~\ref{sec:sim-refpanel} tests directly whether a weaker reference would change the
verdict, and it does not. The original work states a stronger, oracle-conditional guarantee --- that ``our
estimator will always be at least as efficient as an efficient estimator that uses the RCT data only''
\citep{vdlaan2026_atmle}. Our finite-sample $R<1$ characterizes the approach to that limit rather than
contradicting it: the guarantee is the super-efficiency A-TMLE attains when its data-adaptively learned working
model coincides with the oracle bias model, a limit the deployed relaxed-HAL model reaches only as $n$ grows. At
$n_{\mathrm{rct}}=250$ the working model instead pays a finite-sample variance cost that puts $R$ below one
against a matched efficient RCT-only reference. Consistent with the guarantee, the near-zero-bias super-efficiency-like gain does erode toward one as $n$ grows;
under substantial bias, however, $R$ moves \emph{further} below one over the sample sizes studied, because the
deployed relaxed-HAL working model has not yet reached the oracle bias model there --- the asymptotic turnaround
the guarantee predicts lies beyond the studied range (Section~\ref{sec:sim-refpanel}).
\end{enumerate}

The $m=0$ cell is a clean positive control \emph{for this constant-CATE design}: with unbiased RWD, A-TMLE is more efficient
($\sim 1.15\times$), correctly exploiting the extra data (the win is reference- and design-dependent --- it sits at
parity, $\sim 0.92$, against the strengthened matched-glm reference under a heterogeneous within-trial effect; see
the relaxations below and Web Appendix~E). In the lower-to-moderate part of the simulated grid ($m\in[0,2]$) the map reads as ``a modest gain that erodes
toward break-even as the RWD becomes more biased.''

\paragraph{The sub-unity gain is primarily a variance cost in the $1\times$ main grid.}
A variance ratio below one is an \emph{efficiency} verdict only if A-TMLE stays unbiased there; otherwise the
$R<1$ cells would be confounded with mean-squared error from residual bias. Reusing the \emph{same} $B=1000$
replicates without refitting, Table~\ref{tab:biascov} shows the A-TMLE point estimate is approximately unbiased
throughout (relative bias $\le 0.9\%$ through $m=1$, rising only to $1.7\%$/$4.6\%$/$2.5\%$ at the extreme $m=4$
for linear/interaction/wiggly) with 95\% Wald coverage in $0.936$--$0.970$ across all 15 cells and a
well-calibrated SE for $m\le 2$ (mean-SE/MC-SD $\approx 0.96$--$1.03$). Two conclusions follow. First, in the $1\times$ external-data grid the decrease in $R$ is
driven primarily by variance: point-estimate bias is small through $m\le 1$, but becomes visible in the largest and most complex bias cells. Second, a mild,
monotone finite-sample positive bias does emerge at large, \emph{complex} bias (interaction $m=4$: $+0.069$,
$z\approx 9$), the expected regularization residual of the relaxed-HAL working model at $n_{\mathrm{rct}}=250$,
and the one regime where coverage would erode at a larger effect-to-noise ratio.

\begin{table}[htbp]
\centering
\begin{threeparttable}
\caption{\textbf{The sub-unity gain is primarily a variance cost in the 1$\times$ grid.} Reusing the same $B=1000$ replicates
(no refits), each cell reports the A-TMLE bias and, in parentheses, the empirical 95\% Wald-interval coverage.
A-TMLE is approximately unbiased throughout (relative bias $\le 0.9\%$ through $m=1$, rising only to $4.6\%$ at the
extreme interaction $m=4$) with near-nominal coverage, so where the gain of Table~\ref{tab:gainmap} falls below
one the cost is primarily variance: the point-estimate bias stays small through $m=1$ but becomes visible in the largest complex-bias cells.}
\label{tab:biascov}
\small
\begin{tabular}{@{}l ccccc@{}}
\toprule
Bias shape & $m=0$ & $m=0.5$ & $m=1$ & $m=2$ & $m=4$ \\
\midrule
Linear      & $0.000$ (.942) & $-0.008$ (.955) & $-0.002$ (.947) & $0.012$ (.949) & $0.026$ (.970) \\
Interaction & $0.001$ (.942) & $0.005$ (.950)  & $0.011$ (.939)  & $0.031$ (.936) & $0.069$ (.954) \\
Wiggly      & $-0.001$ (.943) & $0.013$ (.945) & $0.009$ (.942)  & $0.019$ (.952) & $0.037$ (.943) \\
\bottomrule
\end{tabular}
\begin{tablenotes}[flushleft]\footnotesize
\item[] Each entry is the A-TMLE point-estimate bias $\widehat\psi-\psi_0$ against the true trial-population ATE
  $\psi_0=1.5$; the value in parentheses is the empirical coverage of nominal 95\% Wald intervals.
\item[] $m$, bias magnitude (bias standard deviation, in outcome-residual SD units). $B=1000$ replicates per cell;
  Monte Carlo SE of coverage $\approx 0.007$.
\end{tablenotes}
\end{threeparttable}
\end{table}

\begin{figure}[htbp]
\centering
\includegraphics[width=0.92\linewidth]{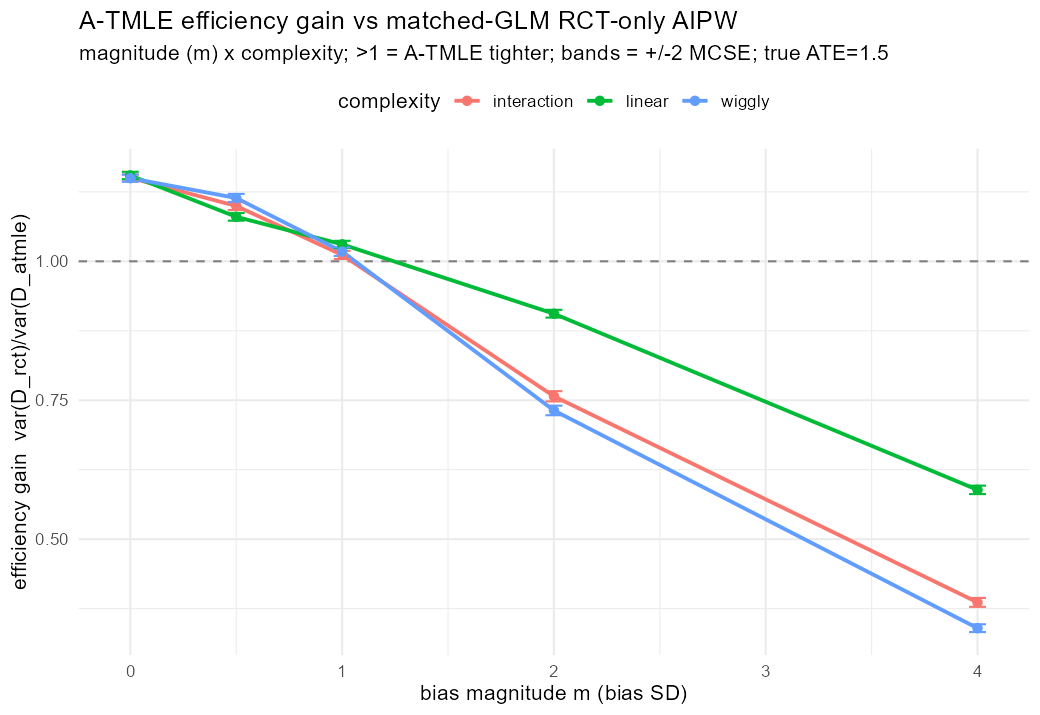}
\caption{\textbf{The efficiency gain map.} Influence-curve gain~\eqref{eq:gain} versus bias magnitude $m$, by
complexity. The gain falls monotonically with magnitude and crosses parity just above $m\approx 1$; complexity
separates the curves only at large magnitude.}
\label{fig:gainmap}

\noindent\emph{Alt text:} Line plot of the efficiency gain (vertical axis) against bias magnitude $m$
(horizontal axis) for three bias complexities. All three curves fall monotonically with $m$ and cross the
value one just above $m=1$, separating from one another only at large $m$ (linear above interaction above
wiggly).
\end{figure}

\paragraph{Two relaxations of the design.} The map above holds trial enrollment completely at random and the
within-trial effect constant. Re-running the full grid after relaxing each in turn (Web Appendix~E; against all
three RCT-only references, $B=1000$ per cell with $B=650$ at $n_{\mathrm{rct}}=400$; the full ladder
$n_{\mathrm{rct}}\in\{250,400,800,1600\}$ for $W$-dependent enrollment and $n_{\mathrm{rct}}\in\{250,400\}$ for the
heterogeneous effect) leaves the
magnitude-over-complexity verdict and the erosion with $n$ intact under both \emph{$W$-dependent trial enrollment}
(realized $\Pi(W)\in[0.10,0.90]$) and a \emph{heterogeneous within-trial effect}
($\mathrm{cate}(W)=1.5+0.8W_1-0.5W_2$, trial ATE still $1.5$). Two qualifications emerge:
\begin{itemize}
\tightlist
\item under selective enrollment A-TMLE's own ATE coverage degrades at the wiggly, large-bias corner, and the
degradation \emph{worsens} with $n$ --- a safe-operating-envelope result we develop in Section~\ref{sec:envelope};
\item under a heterogeneous effect the matched-glm reference becomes strong enough that the gain sits at parity
even at zero bias, while the Super-Learner and HAL references still register a gain --- reinforcing that the gain
is a statement about a chosen reference, not an absolute.
\end{itemize}

\subsubsection{A paper-matched reference panel and the $n$-dependence of the gain}\label{sec:sim-refpanel}
The verdict above is read at $n_{\mathrm{rct}}=250$ against a matched-glm reference. To probe both caveats we
compute the gain map against \emph{three} RCT-only references on the \emph{same} A-TMLE fit ---
matched-glm, a \emph{flexible} discrete SuperLearner $\{$glm, earth, gam$\}$ \citep{vdlaan2007_superlearner}, and relaxed-HAL (the same broad learner family as A-TMLE's default working model, but with
lighter tuning: degree 2 versus the working model's degree 3, both at 5 knots) --- across $n_{\mathrm{rct}}\in\{250,400,800\}$ (total $n$ up to $1600$, the paper's own
scale) and an external multiple $\{1,2,3\}\times$. Every cell is complete with zero reference fall-backs, and
A-TMLE Wald coverage stays in $0.926$--$0.986$. We report both the influence-curve gain~\eqref{eq:gain} and the
original paper's MSE-ratio gain $\mathrm{MSE}(\mathrm{ref})/\mathrm{MSE}(\mathrm{atmle})$; the two track each
other in shape, so the IC-variance gain of Table~\ref{tab:gainmap} is not an artifact of the metric --- though the
MSE ratio runs uniformly a little lower (A-TMLE's small finite-sample bias), so MSE parity is reached marginally
earlier, and a few small-bias cells sit at or just above one on the influence-curve gain yet just below one on the
MSE ratio (Web Appendix~D). Table~\ref{tab:refpanel} summarizes; Web Appendix~D gives the full $n_{\mathrm{rct}}{=}250$ external-multiple
panel and the external-$1\times$ $n$-ladder, together with the MSE-ratio columns; the remaining
$n_{\mathrm{rct}}{=}400$ cross-multiple cells are provided with the reproducibility code.

\begin{table}[htbp]
\centering
\begin{threeparttable}
\caption{\textbf{The sub-unity gain survives a stronger reference and erodes with sample size.}
\emph{Panel A}: influence-curve gain of A-TMLE against three RCT-only references computed on the \emph{same} fit;
a more efficient reference does not raise the gain (the flexible SuperLearner collapses to the GLM on these
near-linear DGPs, and the gain rises only against the noisier relaxed-HAL reference).
\emph{Panel B}: against the efficient (GLM/SL) reference the no-bias gain shrinks toward one and the parity
crossing moves to smaller bias as $n$ grows.}
\label{tab:refpanel}
\small
\begin{tabular}{@{}l cccc@{}}
\multicolumn{5}{@{}l}{\emph{Panel A. Three references} ($n_{\mathrm{rct}}=250$, external multiple $1\times$, Wiggly bias)}\\
\toprule
Reference & $m=0$ & $m=1$ & $m=2$ & $m=4$ \\
\midrule
Matched-GLM (correctly specified)       & 1.156 & 1.015 & 0.729 & 0.344 \\
Flexible SuperLearner (GLM, MARS, GAM)  & 1.169 & 1.024 & 0.735 & 0.347 \\
Relaxed-HAL                             & 1.437 & 1.287 & 0.946 & 0.434 \\
\bottomrule
\addlinespace[6pt]
\multicolumn{5}{@{}l}{\emph{Panel B. Sample-size dependence} (gain vs.\ Matched-GLM, external multiple $1\times$)}\\
\toprule
Sample size & \multicolumn{2}{l}{Gain at $m=0$} & \multicolumn{2}{l}{Crosses 1 near} \\
\midrule
$n_{\mathrm{rct}}=250$\ \ ($n=500$)   & \multicolumn{2}{l}{$\sim 1.15$} & \multicolumn{2}{l}{$m\in(1,2)$} \\
$n_{\mathrm{rct}}=400$\ \ ($n=800$)   & \multicolumn{2}{l}{$\sim 1.09$} & \multicolumn{2}{l}{$m\in(0.5,1)$} \\
$n_{\mathrm{rct}}=800$\ \ ($n=1600$)  & \multicolumn{2}{l}{$\sim 1.05$} & \multicolumn{2}{l}{$m\in(0,0.5)$} \\
\bottomrule
\end{tabular}
\begin{tablenotes}[flushleft]\footnotesize
\item[] Entries are the influence-curve gain $\widehat R=\widehat{\mathrm{var}}(D_{\mathrm{rct}})/\widehat{\mathrm{var}}(D_{\mathrm{atmle}})$;
  $m$ is the bias magnitude. The three references are RCT-only AIPW estimators differing only in the outcome
  regression: a correctly-specified main-terms GLM (the headline denominator), a discrete SuperLearner over
  $\{$GLM, multivariate adaptive regression splines, generalized additive model$\}$, and a relaxed highly-adaptive
  lasso. $B=1000$/$650$/$1000$ across the $n$-ladder ($650$ at $n_{\mathrm{rct}}=400$); A-TMLE Wald coverage $0.926$--$0.986$; zero reference
  fall-backs. GLM, generalized linear model; SL, SuperLearner; HAL, highly adaptive lasso. Full grids and the
  MSE-ratio gains are in Web Appendix~D.
\end{tablenotes}
\end{threeparttable}
\end{table}

\textbf{(1) Not an artifact of a weak denominator.} Against all three references the gain falls below one at large
bias. The flexible references are \emph{not} more efficient than matched-glm: on these near-linear DGPs the
discrete SuperLearner selects the glm fit, so the SL and glm gains agree to within $\sim 1\%$ in most cells (at
most $3.6\%$ at the extreme interaction/high-external corner at $n_{\mathrm{rct}}=250$, shrinking below $1\%$ by
$n_{\mathrm{rct}}=400$), while relaxed-HAL is a \emph{noisier} RCT-only estimator against which A-TMLE looks even
better (Table~\ref{tab:refpanel}, Panel~A). A more efficient reference therefore does not rescue the gain; the
gain rises only against the \emph{noisier} relaxed-HAL, while the genuinely more efficient references (SuperLearner,
matched-glm) leave it essentially unchanged.

\textbf{(2) The finite-sample gain erodes with $n$.} Against the efficient (glm/SL) reference the no-bias gain shrinks
toward one as $n$ grows and the parity crossing moves to smaller bias (Panel~B): from $\sim 1.15$ crossing at
$m\in(1,2)$ at $n=500$, to $\sim 1.05$ crossing at $m\in(0,0.5)$ at $n=1600$. At the paper's own scale A-TMLE is
at or below parity with an efficient RCT-only estimator for all but the smallest bias --- the
finite-sample variance gain is concentrated at small $n$ \emph{and} low bias. This sharpens rather than
overturns the original result.

\textbf{(3) More external data helps only at negligible bias.} Sweeping the external multiple $1\times\to 3\times$
at fixed $n_{\mathrm{rct}}$ (wiggly, MSE-ratio gain), the gain rises only marginally at $m=0$ ($1.06\to 1.08$) and
\emph{falls} once bias is present ($m=1$: $0.89\to 0.82$; $m=2$: $0.61\to 0.43$). The paper's ``more external
data $\Rightarrow$ larger gain'' holds only in the near-unbiased corner; with biased RWD, adding more of it
enlarges the variance A-TMLE must pay to correct. Consistent with this, under the $3\times$-external arm A-TMLE's
worst-case relative bias rises to $\sim 9\%$ (interaction, $m=4$, $n_{\mathrm{rct}}=250$: $9.2\%$; $n=400$:
$8.8\%$), still with near-nominal coverage ($0.93$) --- a caveat to the ``pure variance cost'' reading that is
invisible in the $1\times$ map.

\textbf{Coverage is robust to the HAL smoothing choice under random enrollment.} A dedicated undersmoothing slice (penalty multiplicity
$n_\lambda\in\{1,3,5\}$ at the highest-bias cells) confirms the $m=4$ finite-sample bias does not erode validity:
A-TMLE's 95\% Wald coverage stays near-nominal ($0.93$--$0.96$) across the undersmoothing range at both
$n_{\mathrm{rct}}=250$ and $400$, with only a small monotone bias reduction at wiggly $m=4$ ($3.4\%\to 1.8\%$ at
$n=250$).

\subsection{Why magnitude, not complexity, drives the variance}\label{sec:theory}
The first headline finding of the efficiency map --- that the gain is governed by bias \emph{magnitude} rather than
complexity --- reflects a structural feature of the pooled-projection variance rather than a quirk of the chosen
bias shapes. The efficiency loss is localized in $\mathrm{var}(D_A)$, the
variance of the pooled-projection influence curve, and admits an exact population-oracle analytic identity
(Proposition~\ref{prop:varDA}) under an explicitly restricted oracle model. To our knowledge this
magnitude-dominance identity is a new result; its full derivation is given in Web Appendix~G. (The companion
question --- why the naive and cross-fit standard errors fail --- is taken up where those failures appear, in
Section~\ref{sec:sim-selse}.)

The identity uses four objects from the influence-curve decomposition of Web Appendix~A: the working-model basis
$\Phi$ that the pooled CATE $\tau_A$ is fit on; the outcome-noise variance $\sigma^2=E[U_Y^2\mid W,A]$; the pooled
$R$-learner residual $r=Y-\theta-(A-g)\tau_A$ with $\theta=E(Y\mid W)$; and $D_A^W$, the covariate (heterogeneity)
component of the pooled-projection influence curve $D_A=D_A^W+D_A^\beta$.

\begin{proposition}[population curvature of the oracle, intercept-only influence curve]\label{prop:varDA}
Under the main design, oracle nuisances, and a \emph{forced intercept-only working model} ($\Phi\equiv1$) --- an
assumed restriction, since a homogeneous within-trial effect does \emph{not} by itself make the pooled projection
$\tau_A(W)$ constant or collapse the HAL selection to the intercept once the external arm carries an arm-dependent
bias --- the population variance of the (oracle) pooled-projection influence curve $D_A$ is the exact identity
\begin{equation}
\mathrm{var}(D_A) = a + b\,m^2,
\quad
a=\mathrm{cc}\,\sigma^2,\quad
b=\mathrm{cc}^2\,\frac{E\big[(A-g)^2\,\widetilde s^{\,2}\big]}{\mathrm{sd}_0(s)^2}\ge0,
\label{eq:prop1}
\end{equation}
where $\mathrm{cc}=1/E\{g(1-g)\}$ (the pooled inverse-overlap constant), $g$ is the pooled propensity, and $\widetilde s$ is the $L^2$ residual of the
unit-bias carrier $\mathbf1\{S{=}0\}s(W,A)$ after removing its $W$-conditional mean and its $(A-g)$-component (the
explicit formula and the shape/$W$ specifications that make $a,b$ reproducible are in Web Appendix~G). There is no
linear-in-$m$ term. The identity is population-oracle exact --- not a finite-sample identity for the deployed
HAL-selected estimator.
\end{proposition}

\noindent
The bias \emph{magnitude} thus enters at leading order, quadratically through $m^2$, with no linear-in-$m$ term and no shape dependence at this order.
The mechanism is transparent: the forced intercept-only model leaves only the score
term $\mathrm{cc}\,(A-g)\,r$; the pooled outcome regression and pooled CATE remove the $W$-measurable and
$(A-g)$-components of the injected bias, and the residual carrier contributes a variance \emph{exactly} proportional
to $m^2$ (the linear term vanishes identically because the outcome noise is conditionally mean-zero, $E[U_Y\mid W,A,S]=0$). The curvature
$b$ is an \emph{arm-weighted} second moment: because the planted bias depends on the treatment arm, the symmetric
$g(1-g)$ factorization fails and the correct expression carries arm-asymmetric weights. In the full influence curve $D=D_A-D_S$, three couplings could in principle bring bias shape, complexity, or noise
into the $\mathrm{var}(D_A)$ identity above at first order; none does.
\begin{enumerate}
\item \emph{Heterogeneity$\,\times\,$bias.} It does not enter, because the forced intercept-only working model sets $D_A^W\equiv0$, removing the pooled projection's covariate (heterogeneity) component from $D_A$.
\item \emph{Membership-score$\,\times\,$bias.} It does not enter, because the trial-membership score (the $\Pi^\star$-fluctuation score) is carried by the bias-correction term $D_S$, not by $D_A$. In $\mathrm{var}(D)=\mathrm{var}(D_A)-2\,\mathrm{Cov}(D_A,D_S)+\mathrm{var}(D_S)$ any such pairing sits inside $\mathrm{var}(D_S)$ or the cross-covariance $\mathrm{Cov}(D_A,D_S)$, never inside $\mathrm{var}(D_A)$; it is absorbed there, not a free $D_A$--$D_S$ cross term, and we do not certify its magnitude.
\item \emph{Bias--noise cross term.} The one such term internal to $\mathrm{var}(D_A)$ --- outcome noise $U_Y$ times the centered bias carrier --- drops by $E[U_Y\mid W,A,S]=0$ (Web Appendix~G, Step~4), a step that uses conditional mean-zero alone, neither $S\perp(A,W)$ nor constancy of $\Pi$. (Indeed $\Pi(W)$ is not even the relevant object for $D_S$, whose targeted score runs through the arm-dependent $\Pi(W,A)$, so constancy of $\Pi(W)$ is not the reason here.)
\end{enumerate}
Once the working model is richer than the intercept ($\Phi\not\equiv1$, restoring $D_A^W$) or enrollment is $W$-dependent, these couplings re-enter the variance budget along \emph{different} routes. The heterogeneity$\,\times\,$bias coupling then enters $\mathrm{var}(D_A)$ directly through $\mathrm{var}(D_A^W)$, while the membership-score coupling enters through $\mathrm{var}(D_S)$ and the cross term $\mathrm{Cov}(D_A,D_S)$. We treat all of these empirically rather than as a theorem (Web Appendices~E, G).

Computing the population curvature directly gives per-shape $b=\{1.49,1.38,1.16\}$ (linear, interaction, wiggly),
mean $\approx1.34$. The arm-weighting flagged above is not a technicality. A natural but incorrect route to $b$
makes three substitutions at once: it factors $E[(A-g)^2\widetilde s^{\,2}\mid W]=g(1-g)\,E[\widetilde s^{\,2}\mid W]$
--- valid only if the carrier were arm-free --- uses the within-$W$ arm variance in place of the centered carrier's
second moment, and drops the $\theta$ and $\tau_A$ projections that define $\widetilde s$. Together these give
$b=\{0.43,0.35,0.18\}$, understating the correct curvature three- to seven-fold. The arm-weighting is the first of
the three: the correct weights are asymmetric, $g(1-g)^2$ on the treated arm and $(1-g)g^2$ on the control arm
(Web Appendix~G, Step~5, where the three defects are separated).
A simulated corroboration regresses per-cell $\mathrm{var}(D_A)$ on $m^2$ and the selected-basis
count $\bar d$ across the $15$-cell map (Web Appendix~D): $\mathrm{var}(D_A)\approx 3.25 + 1.40\,m^2 + 0.094\,\bar d$
($R^2=0.989$); the population constant $a=\mathrm{cc}\,\sigma^2=4.18$ reproduces the empirical $m{=}0$ value of $\mathrm{var}(D_A)$
to three digits (the regression's own intercept $3.25$ is lower because the $0.094\,\bar d$ term carries part of
the $m{=}0$ level),
the mean curvature ($\approx1.34$) sits just below the fitted $1.40$, and the standardized magnitude effect ($0.91$)
is eight times the complexity effect ($0.11$) --- though, because $\bar d$ is endogenous and collinear with $m^2$,
we read this standardized comparison as indicative of magnitude-dominance rather than a clean variance decomposition. One subtlety: the \emph{population} shape ordering
(linear$\,>\,$interaction$\,>\,$wiggly --- a wigglier bias projects more into the absorbed span) is the \emph{reverse}
of the empirical finite-$n$ ordering, an empirical phenomenon driven by the finite-$n$ channels the Proposition
excludes (the $\bar d$/selector inflation and finite-HAL nuisance-fit error); the $\bar d$ term is thus an empirical
finite-sample extension, not part of Equation~\eqref{eq:prop1}. The parity crossing involves the combined
$\mathrm{var}(D_A-D_S)$ and is read near $m\approx1$ residual SD empirically.

\emph{Takeaway:} the population identity $\mathrm{var}(D_A)=a+b\,m^2$ makes bias \emph{magnitude} the leading-order driver of the variance, with curvature $b\approx1.34$ and a standardized magnitude effect ($0.91$) about eight times the complexity effect ($0.11$). Within its restricted scope --- population, oracle nuisances, forced intercept-only working model --- this gives a tractable population characterization of why bias magnitude drives $\mathrm{var}(D_A)$: it is not the complexity-driven oracle super-efficiency mechanism, it does not predict the empirical shape ordering (which the finite-$n$ channels reverse), and it is not a finite-sample statement about the deployed HAL-selected estimator, whose empirical behaviour we track separately.

\subsection{An empirical safe operating envelope for A-TMLE's own inference}\label{sec:envelope}
The main design enrolls units into the trial completely at random, so the membership probability $\Pi(W)$ is
constant and the trial-enrollment positivity assumption of Section~\ref{sec:setup} holds vacuously. To make that
assumption nontrivial we replace random enrollment with a \emph{selective} rule, drawing
$S\mid W\sim\mathrm{Bernoulli}(\Pi(W))$ with a $W$-dependent logit, and scan
$n_{\mathrm{rct}}\in\{250,400,800,1600\}$ at $\mathrm{ext}=1$ so that $\Pi(W)$ is the identical function of $W$ at
every rung. The selection is deliberately strong: the central $90\%$ of $\Pi(W)$ falls in $[0.10,0.90]$, but
$1.4\%$ of the covariate distribution lies outside $[0.05,0.95]$, so the expected number of near-violating units
grows with the sample ($\approx7$ at $n_{\mathrm{total}}=500$, $\approx44$ at $3{,}200$) and realized minima reach
$\approx0.015$. Trial-enrollment positivity is therefore \emph{stressed} by design rather than comfortably
satisfied; the stress deepens with $n$ alongside the coverage loss. It is not,
however, a sufficient explanation for what follows: the interaction and the two rough packet surfaces below run
under the \emph{identical} $\Pi(W)$ at the identical $n$ and keep $0.94$--$0.96$ coverage, so near-positivity alone
does not produce the effect. The trial ATE stays identified and equal to $1.5$ throughout --- within-trial
randomization and a homogeneous within-trial effect fix it regardless of who enrolls. Here we track a different quantity from the rest of
Section~\ref{sec:sim}: A-TMLE's \emph{own} $95\%$ Wald coverage of that ATE, not the efficiency gain. The primary
analysis fits the nuisances ($\theta$, $Q$, $\Pi$, $g$) by main-terms GLM, as throughout; a Super-Learner
comparison enters where we diagnose the mechanism below.

\begin{table}[htbp]
\centering
\begin{threeparttable}
\caption{\textbf{A safe operating envelope: A-TMLE's own ATE coverage under $W$-dependent enrollment.} Wald
coverage of the true trial ATE ($=1.5$), the study's baseline $5$-knot HAL tuning, $\mathrm{ext}=1$ so $\Pi(W)$ is the identical
function of $W$ at every sample size. Under main-terms GLM nuisances, coverage falls with $n$ for one bias surface
only, and a Super-Learner nuisance fit on the identical data restores it across the ladder (Panel~A). The coverage
deterioration appears only under selective enrollment in this comparison; the analysis does not establish selective
enrollment as a sufficient cause (Panel~B). The working-model tuning knobs do not restore it,
including a penalty path driven to the undersmoothing criterion (Panel~C). Targeting drift, not the basis count, is
what an analyst can see without ground truth (Panel~D).}
\label{tab:envelope}
\small
\begin{tabular}{@{}l cccc c@{}}
\multicolumn{6}{@{}l}{\emph{Panel A. Coverage by bias surface and total sample size ($m=4$, selective enrollment).}}\\
\toprule
$n_{\mathrm{total}}$          & $500$  & $800$  & $1{,}600$ & $3{,}200$ & slope on $\log n$\tnote{a} \\
\midrule
Linear (smooth)               & $0.96$ & $0.96$ & $0.95$ & $0.91$ & $-0.50$ \\
Interaction                   & $0.94$ & $0.94$ & $0.94$ & $0.95$ & $+0.10$ \\
Rough (Gabor packet in $W_2$) & ---    & ---    & $0.96$ & $0.96$ & $+0.14$ \\
Rough (same packet in $W_1$)  & ---    & ---    & $0.94$ & $0.96$ & $+0.59$ \\
\textbf{Wiggly (rough), GLM nuisances}       & $\mathbf{0.81}$ & $\mathbf{0.73}$ & $\mathbf{0.64}$ & $\mathbf{0.47}$ & $\mathbf{-0.82}$ \\
\quad same surface, SL nuisances & $0.91$ & $0.91$ & $0.94$ & $0.94$ & $+0.31$ \\
\bottomrule
\end{tabular}

\vspace{0.7em}
\begin{tabular}{@{}l ccc@{}}
\multicolumn{4}{@{}l}{\emph{Panel B. Same wiggly surface, varying only enrollment selectivity ($m=4$, $n_{\mathrm{total}}=3{,}200$).}}\\
\toprule
Enrollment                          & random   & intermediate  & selective \\
$\Pi(W)$ by design                  & constant & $W$-dependent & strongly $W$-dependent \\
Fitted $\widehat\Pi(A,W)$ spread\tnote{b} & $0.22$ & $0.55$ & $0.82$ \\
\midrule
A-TMLE coverage                     & $0.97$   & $0.89$   & $0.47$ \\
Relative bias                       & $+0.6\%$ & $-5.7\%$ & $-13.1\%$ \\
\bottomrule
\end{tabular}

\vspace{0.7em}
\begin{tabular}{@{}l ccc c ccc@{}}
\multicolumn{8}{@{}l}{\emph{Panel C. Fixed-$n$ tuning at the wiggly corner ($m=4$, $n_{\mathrm{total}}=1{,}600$).}}\\
\toprule
                 & \multicolumn{3}{c}{HAL knots (grow the sieve)} & & \multicolumn{3}{c}{penalty $n_\lambda$ (relax)} \\
\cmidrule(lr){2-4}\cmidrule(lr){6-8}
                 & $5$    & $10$   & $20$   & & $1$    & $3$    & $5$    \\
\midrule
A-TMLE coverage  & $0.64$ & $0.50$ & $0.50$ & & $0.64$ & $0.59$ & $0.70$ \\
\bottomrule
\end{tabular}

\vspace{0.7em}
\begin{tabular}{@{}l ccccc@{}}
\multicolumn{6}{@{}l}{\emph{Panel D. Ground-truth-free observables at $m=4$, $n_{\mathrm{total}}=3{,}200$, selective enrollment.}}\\
\toprule
                         & Linear & Interaction & Rough ($W_2$) & Rough ($W_1$) & \textbf{Wiggly} \\
\midrule
Active bases             & $18.9$ & $50.9$  & $56.1$  & $64.8$  & $61.4$ \\
Targeting drift\tnote{c} & $0.019$ & $0.011$ & $0.008$ & $0.011$ & $\mathbf{0.057}$ \\
\midrule
A-TMLE coverage          & $0.91$ & $0.95$  & $0.96$  & $0.96$  & $\mathbf{0.47}$ \\
\bottomrule
\end{tabular}
\begin{tablenotes}[flushleft]\footnotesize
\item[a] Replicate-clustered logistic slope of coverage on $\log n_{\mathrm{total}}$. Only the wiggly slope is both
  large and highly significant ($p<10^{-80}$); the linear slope is significant but leaves coverage $\ge0.91$
  throughout, and the three remaining shapes are flat.
\item[b] Spread (P5--P95) of the \emph{fitted} $\widehat\Pi(A,W)$; the $0.22$ under random enrollment is estimation
  noise plus the arm dependence of $\Pi(A,W)$, shown to confirm the fit separates the three designs.
\item[c] Root-mean-square movement of the learned enrollment-effect surface under targeting, weighted by
  $\widehat\Pi(1-\widehat\Pi)$; unnormalized, on the outcome scale. Section~\ref{sec:envelope} reports its pooled
  discrimination, the stability-of-ordering argument the paper rests on, and why a scale-normalized drift is worse.
\item[] Coverage over $B=400$--$2{,}400$ replicates per cell (pooled across rungs that share a configuration and
  draw independent seeds). Across Panel~A's GLM wiggly row the relative bias is flat ($-12.8$ to $-13.1\%$) while the
  Monte Carlo SD shrinks $\approx$ two-thirds and $\mathrm{mean\text{-}SE}/\mathrm{MC\text{-}SD}$ moves
  $0.84\to1.02$: the loss is a persistent \emph{centering} bias, not standard-error under-estimation.
\end{tablenotes}
\end{threeparttable}
\end{table}

\paragraph{Coverage falls below nominal for one surface, and the shortfall deepens with $n$.} A-TMLE's interval is
well calibrated for most enrollment-effect surfaces and stays so under selective enrollment. It falls below nominal
for the \emph{wiggly} surface at large bias, and there the shortfall deepens as the sample grows
(Table~\ref{tab:envelope}, Panel~A): at $m=4$, coverage falls monotonically from $0.81$ at
$n_{\mathrm{total}}=500$ to $0.47$ at $3{,}200$, a replicate-clustered logistic slope on $\log n$ of $-0.82$
($p<10^{-80}$). The degradation is graded rather than a cliff: at $n_{\mathrm{total}}=3{,}200$ coverage runs
$0.93,0.90,0.83,0.69,0.47$ across $m=0,0.5,1,2,4$, with the shortfall becoming material near $m\approx1$ residual
SD --- the same magnitude at which the efficiency gain crosses parity (Section~\ref{sec:sim-map}).

Four comparison surfaces stay near nominal, and together they narrow what can be blamed. The interaction shape is
likewise beyond a main-terms GLM's reach yet keeps $\approx0.94$--$0.95$, so \emph{generic} nuisance
misspecification is not the cause. More pointedly, two further surfaces that are genuinely rough --- a Gabor packet
$\exp(-W^2/2)\cos(3.5W)$, placed once on $W_2$ and once on $W_1$ --- also keep nominal coverage with flat slopes
($+0.14$, $+0.59$), even though they carry basis counts as large as the wiggly surface's. Roughness alone is
therefore not the operative feature, and neither is placing the roughness on the covariate that carries the
enrollment score's own nonlinearity: the $W_1$ packet differs from the $W_2$ packet in exactly that respect and
behaves identically. What distinguishes the wiggly surface from both packets is that its oscillation does not decay
and it carries an unbounded quadratic term, so its bias contribution persists into the tails of $W_1$ --- precisely
where selective enrollment drives $\Pi(W)$ toward its bounds. That reading is only partially consistent with the
five surfaces, and we say so: the linear surface, whose bias also grows without bound in those tails, is the only
other one with a significantly negative slope ($-0.50$, though it stays $\ge0.91$ throughout), but the interaction
surface contains the same linear term and is flat. So tail growth is at best a contributing feature, not a
sufficient one. We report the pattern rather than a mechanism; isolating it would need surfaces designed to vary
tail behaviour while holding everything else fixed, which the surfaces we ran do not.

\paragraph{The mechanism is a persistent centering bias.} Across the ladder the relative bias of A-TMLE's ATE
stays essentially constant ($\approx-13\%$) while the Monte Carlo standard deviation shrinks by roughly two-thirds
($0.26\!\to\!0.09$, close to root-$n$). The model standard error is mildly optimistic at small $n$
(mean-SE$/$MC-SD $=0.83$--$0.84$ at $n_{\mathrm{total}}\le800$) but well calibrated at the large-$n$ corner
where coverage falls hardest ($0.96$--$1.02$ at $n_{\mathrm{total}}\ge1{,}600$). The with-$n$ collapse is
therefore, unambiguously, a \emph{centering} phenomenon --- the interval tightens around a biased centre ---
exactly where the effect is worst. The binding constraint is the outcome-nuisance fit, not the working-model
dictionary. Holding the dictionary fixed and replacing the main-terms GLM nuisances with a Super Learner ensemble
($\{$GLM, MARS, GAM, HAL$\}$ under a convex meta-learner), on the identical datasets and cross-validation folds,
restores near-nominal coverage across the whole ladder --- $0.91,0.91,0.94,0.94$ at
$n_{\mathrm{total}}=500,800,1{,}600,3{,}200$, against the GLM row's $0.81,0.73,0.64,0.47$ --- and drives the
relative bias to zero ($-4.1\%$ at $n_{\mathrm{total}}=500$ to $-0.3\%$ at $3{,}200$) rather than leaving it flat
(Table~\ref{tab:envelope}, Panel~A; the full per-cell grid for both arms is in Web Appendix~E). The same fixed dictionary therefore represents the surface well once the
nuisances improve, so the earlier ``the dictionary cannot span the surface'' reading is not the mechanism. The
route is direct: the enrollment-effect working model is fit by an $R$-learner whose pseudo-outcome
$(Y-\widehat\theta)/(A-\widehat g)$ carries the nuisance residual, so a misspecified $\widehat\theta$ propagates
into the very target the working model must reproduce. The persistent GLM bias is thus the finite-sample
working-model approximation error of Section~\ref{sec:setup} made first-order \emph{through} the outcome-nuisance
channel, not by a dictionary too coarse to resolve the surface.

\paragraph{Coverage deterioration appears only under selective enrollment in this comparison.} Holding the wiggly surface, the tuning, and
the sample size fixed and varying \emph{only} how selective enrollment is settles the attribution
(Table~\ref{tab:envelope}, Panel~B). Under random enrollment --- the identical rough surface, with $\Pi(W)$
constant --- coverage at $n_{\mathrm{total}}=3{,}200$ is $0.97$ and the relative bias is $+0.6\%$; across the ladder
coverage there \emph{improves} with $n$ (slope $+0.43$, $95\%$ CI $[+0.14,+0.72]$), as it should. At intermediate
selectivity coverage is $0.89$ with a $-5.7\%$ bias, and at full selectivity $0.47$ with $-13.1\%$. The
degradation is thus monotone in enrollment selectivity and absent without it. The envelope is a statement about
trial-enrollment positivity being stressed, not about rough outcome surfaces \emph{per se}.

\paragraph{Neither the tuning we varied nor undersmoothing driven to its criterion restores coverage.} The primary evidence against a tuning fix is Panel~A
itself: under fixed default tuning the lasso-selected basis already grows $25\!\to\!61$ as $n$ increases, yet
coverage falls to $0.47$. Panel~C varies two tuning knobs directly, at a single $n_{\mathrm{total}}=1{,}600$.
Enriching the HAL sieve from $5$ to $10$ to $20$ knots drops coverage to $\approx0.50$ and it does not recover: at
fixed $n$ a richer basis overfits the selection-stressed structure. This is a finite-$n$ gap between the
asymptotically-motivated undersmoothing prescription --- whose remedy grows resolution \emph{with} $n$ --- and the
sample sizes deployment sees, not a defect of that prescription; a formal $n$-indexed undersmoothing schedule is
untested here and is the natural next check. Relaxing the lasso penalty ($n_\lambda=1,3,5$) nudges coverage from
$0.64$ toward $0.70$, in the right direction but still far below nominal at this corner.

We are explicit about the limit of this evidence, which the Targeted Learning literature would naturally scrutinize. Neither knob was driven by the criterion that defines undersmoothed HAL --- decreasing $\lambda$ until
the empirical scores generated by the basis functions are controlled at $\widehat\sigma/(\sqrt{n}\log n)$. At the
default tuning we instrumented that criterion directly and it is not met, by a factor of a few; the depth it
demands is on the order of tens of additional steps down the penalty path, against the one, three, and five steps
Panel~C varies. So we drove the path to the criterion directly, at the worst corner ($n_{\mathrm{total}}=3{,}200$,
wiggly, $m=4$, on a moderately enriched ten-knot sieve): decreasing $\lambda$ per replicate until the empirical
basis scores are controlled at $\widehat\sigma/(\sqrt n\log n)$. The criterion is then met in every replicate
(mean score ratio $0.955\le1$) at a depth of about fourteen steps below $\lambda_{\min}$ --- and coverage still
reaches only $0.62$, against the $0.94$ that flexible nuisances deliver at the identical corner. Criterion-driven
undersmoothing thus moves the interval in the right direction but does not repair it, whereas the nuisance change
does. We do not over-read a single sample size: a formal $n$-indexed undersmoothing schedule, whose resolution
grows \emph{with} $n$, remains the natural next check rather than a settled negative. We report the criterion in
its scale-free form, since with first-order spline bases the raw-coordinate version is not invariant to the units
the covariates happen to be recorded in.

\paragraph{What separates the regimes in the simulation.} The boundary is visible from the report card, but not
through the quantity one might reach for first. At matched $m$ and $n$ the active-basis count does not discriminate
\emph{reliably}. Ranking the $m=4$ surfaces by mean active-basis count does put the failing one first at
$n_{\mathrm{total}}=500$, $800$ and $1{,}600$ --- but the ordering inverts at $3{,}200$, where the $W_1$ packet
carries $64.8$ bases at $0.96$ coverage against the wiggly surface's $61.4$ at $0.47$
(Table~\ref{tab:envelope}, Panel~D). A quantity that ranks correctly until the sample is large enough for the
problem to be worst is not a usable flag. The \emph{targeting drift} ranks the failing surface first at every
sample size ($0.031$, $0.034$, $0.048$, $0.057$), and its margin widens with $n$ rather than closing: at
$3{,}200$ it separates by five- to sevenfold with no overlap, placing the linear surface ($0.019$, coverage
$0.91$) between the two groups. Pooling all $73$ cells, drift attains AUC $0.90$ for identifying cells with
coverage below $0.90$, against $0.85$ for the basis count --- but that gap is not itself established, and we do
not claim it: the two are within sampling error of each other, and the cells are not independent, sharing rungs,
seeds and nested sample sizes. What the evidence supports is the ordering being stable for drift and not for the
basis count, not a calibrated detector. The drift here is the unnormalized quantity; dividing it by the learned
surface's own scale makes it worse, because that scale is near zero in the no-bias cells.

One caution, because it inverts within a setting. \emph{Across} configurations, high drift marks the stressed
regime. \emph{Within} the collapsing cell, the replicates that miss are those where the lasso selected the
\emph{sparser} working model --- at $n_{\mathrm{total}}=1{,}600$, missed intervals average $39.8$ bases and drift
$0.038$ against $49.7$ and $0.053$ for the covered ones --- because a sparse correction under-corrects. A large
drift should therefore be read as a property of the regime, and a small one on a single dataset inside that regime
is not reassurance.

\emph{Takeaway:} A-TMLE's own interval is trustworthy across the efficiency map we studied --- for smooth and moderately
complex enrollment-effect surfaces, for genuinely rough ones, at non-extreme bias, and under selective enrollment.
It marks one outer corner where it is not: a bias surface whose structure persists where selective enrollment
drives $\Pi(W)$ toward its bounds, at large bias magnitude, where the shortfall deepens with $n$. There the
binding constraint is the outcome-nuisance fit: main-terms GLM nuisances leave a persistent centring bias that
neither the tuning we varied nor undersmoothing driven to its criterion repairs, whereas a Super Learner ensemble
restores near-nominal coverage across the ladder --- the extension the Targeted Learning program prescribes.
Selective enrollment is necessary for the effect --- it vanishes under random enrollment --- and the report card's
targeting drift, not its basis count, is what makes the corner visible from observables alone. Charting that boundary is part of empirically establishing where the estimator can be trusted, and the trial-only estimate should
lead inside it.

\subsection{Selection-aware inference for the efficiency gain}\label{sec:sim-selse}
Having accounted for the magnitude-dominance of the variance, we turn to the second headline finding: how to put a
selection-aware interval on the gain. Across the 15-cell sweep the naive SE is anti-conservative: $\mathrm{mean(SE)}/\mathrm{MC\text{-}SD}\approx
0.53$--$0.74$, with fixed-truth coverage $0.70$--$0.87$, worst at large $m$ in this sweep; the undercoverage is not,
however, monotone in model size --- the naive SE's single worst cell is instead the large-$n$ \emph{no-bias}
corner (Web Appendix~E), where the selected model is smallest. We therefore ask which of the ten selection-aware SEs of Section~\ref{sec:methods-selse} is
calibrated, on $7{,}500$ datasets (the 15 cells, $B=500$). \textbf{Coverage must be scored against a fixed truth}
--- the locked $B=1000$ gain of Table~\ref{tab:gainmap} --- not against each method's own mean, which would grade
a biased estimator against a reference that moves with the estimator. The gap between the two scorings is itself diagnostic: the cross-fit,
CV-TMLE, and HulC comparators look respectable self-centered yet undercover markedly against the fixed truth.
That fixed truth is close to the naive gain's Monte-Carlo expectation, so the naive and block-jackknife estimators
--- which share the full-data point estimate --- are graded against essentially their own centre; the clean,
decision-relevant contrast is therefore \emph{naive versus jackknife}, a test of interval \emph{width} at a common
centre that the naive SE fails despite being correctly centred (its coverage stays $0.70$--$0.87$, falling to
$\approx0.28$ in the large-$n$ no-bias cells). The cross-fit/CV-TMLE shortfall against the fixed truth instead
combines a biased centre with a genuinely mis-calibrated SE --- the latter shown by their \emph{own-mean} coverage
also falling below nominal (cross-fit $0.60$--$0.90$).
Table~\ref{tab:selse} summarizes.

\begin{table}[htbp]
\centering
\begin{threeparttable}
\caption{\textbf{Ten selection-aware standard errors for the efficiency gain, scored against a fixed truth.}
Each method estimates the gain and a standard error of its logarithm; entries are ranges over the 15 cells
(Linear/Interaction/Wiggly $\times\ m\in\{0,0.5,1,2,4\}$), $B=500$. Only the block jackknife attains nominal
(markedly conservative) coverage; no other method reaches $0.93$ in any cell. The gap between the fixed-truth and
own-mean coverage columns is itself diagnostic: the cross-fit, CV-TMLE, and HulC comparators look
respectable self-centered yet undercover markedly against the fixed truth. Per-method compute time (all ten) is reported in
Web Appendix~F: every fold-based SE costs $\approx V$ A-TMLE fits, so the block jackknife is no more expensive than
the variants that undercover.}
\label{tab:selse}
\small
\begin{tabular}{@{}l cccc@{}}
\toprule
Method & Gain bias\tnote{a} & Cover., fixed\tnote{b} & Cover., own mean\tnote{c} & Ratio\tnote{d} \\
\midrule
\textbf{Block jackknife} & $-0.8$ to $+2.1\%$ & \textbf{0.984--0.998} & 0.984--0.998 & 1.6--2.2 \\
Naive IF-SE & $-0.8$ to $+2.1\%$ & 0.704--0.866 & 0.71--0.86 & 0.53--0.74 \\
Cross-fit (plain out-of-fold) & $-8$ to $-65\%$ & 0.15--0.52 & 0.60--0.90 & 0.24--0.57 \\
\quad $+$ ridge $\mathrm{IM}^{-1}$ ($\lambda{=}0.01/0.05/0.2$) & $-41$ to $+130\%$ & 0.00--0.84 & 0.53--0.82 & 0.31--0.62 \\
\quad $+$ winsorized IC & $-10$ to $-40\%$ & 0.12--0.44 & 0.53--0.77 & 0.33--0.59 \\
\textbf{CV-TMLE} (re-target per fold) & $-8$ to $-56\%$ & 0.18--0.51 & 0.67--0.88 & 0.43--0.74 \\
\quad CV-TMLE $+$ ridge $\mathrm{IM}^{-1}$ & $-27$ to $+78\%$ & 0.03--0.78 & 0.56--0.81 & 0.42--0.62 \\
HulC (assumption-lean hull) & $+251$ to $+461\%$ & 0.07--0.57 & 0.84--0.90 & 1.0--1.2 \\
\bottomrule
\end{tabular}
\begin{tablenotes}[flushleft]\footnotesize
\item[a] Relative bias of the method's mean gain against the locked $B=1000$ truth (Table~\ref{tab:gainmap}).
\item[b] Coverage of the 95\% CI against that \emph{fixed} truth --- the common reference value.
\item[c] Coverage against the method's own mean (circular; shown to expose the gap).
\item[d] Calibration ratio mean(SE)$/$Monte-Carlo SD of $\log(\text{gain})$; $1$ is calibrated, the jackknife is
  conservative. IF, influence function; IM, information matrix; CV-TMLE, cross-validated targeted maximum
  likelihood estimation; HulC, the convex-hull interval of \citet{kuchibhotla2024_hulc}. Monte Carlo SE of
  coverage $\le 0.022$.
\end{tablenotes}
\end{threeparttable}
\end{table}

\begin{figure}[htbp]
\centering
\includegraphics[width=0.92\linewidth]{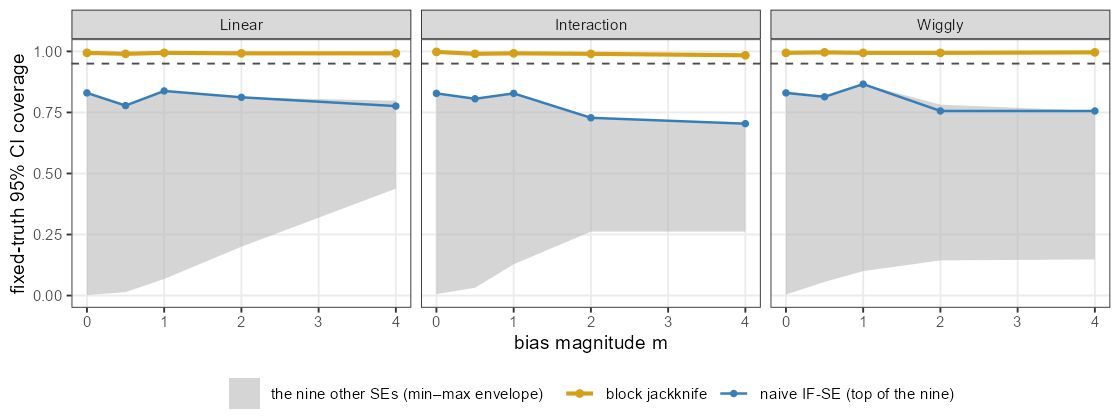}
\caption{\textbf{Fixed-truth coverage of the ten selection-aware SEs.} 95\% CI coverage of the efficiency gain,
scored against the locked $B=1000$ truth, by bias magnitude $m$ and shape (faceted); the dashed line marks the
$0.95$ target. The block jackknife (gold) is the only method to reach the target ($0.98$--$1.00$); the other nine fall entirely
within the shaded min--max envelope (never above $0.87$), undercovering across all three shapes, with naive (blue)
the best of them at $0.70$--$0.87$. The full per-method breakdown is in Table~\ref{tab:selse} and Web Appendix~D.}
\label{fig:selsecov}

\noindent\emph{Alt text:} Faceted line plot of 95\% confidence-interval coverage of the efficiency gain
(vertical axis) against bias magnitude (horizontal axis), one facet per bias shape. The block-jackknife curve
is the only one to reach the 0.95 target line (0.98--1.00); the other nine methods stay within a shaded
envelope that never exceeds 0.87.
\end{figure}

\paragraph{The block jackknife gives conservative empirical coverage.}
The block jackknife --- delete-a-fold, re-selecting the working model on each $S$-stratified leave-fold-out
subsample --- is the only method that is both approximately unbiased ($\le 2.1\%$, inherited from the naive
full-data gain it centers on) and attains nominal coverage ($0.984$--$0.998$ across the main 15 cells, markedly
conservative; the ratio $\approx 1.6$--$2.2$ overshoot reflects right-skew and the non-smooth, selection-based
nature of the statistic, for which we do not establish consistency or asymptotic exactness). It is the only one of the ten whose fixed-truth
coverage reaches $0.95$ --- indeed no other method reaches even $0.93$ in any of the 15 cells.

\paragraph{The cross-fit family did not calibrate here.}
The cross-fitted out-of-fold variance was the theoretically-motivated refinement, and the specific reconstruction we implemented did not deliver it. Its fixed-truth coverage is $0.15$--$0.52$, which the self-centered $0.60$--$0.88$ had masked, because its gain \emph{point estimate} is biased. Cross-fit underestimates the gain by up to $65\%$ and the ridge variants swing from $-41\%$ to $+130\%$, so these are poor estimators of the gain, not merely mis-calibrated SEs. The downward bias comes from an uncentered out-of-fold reconstruction, a finite-sample effect that shrinks with $n$; the closing heuristic of this subsection makes the mechanism precise. Ridge and winsorization shift it around but do not restore coverage in our experiments (still $0.00$--$0.84$). The full ten-method, per-cell results are in Web Appendix~D.

\paragraph{The principled CV-TMLE also did not calibrate here.}
The comparator the theory points to --- a CV-TMLE that re-targets the working-model fluctuation on each held-out fold (one fluctuation each for $\tau_A$ and $\tau_S$, centering the validation score plain cross-fit leaves uncentered, in the spirit of the original paper's Appendix-C construction) --- also did not calibrate in the studied design: fixed-truth coverage $0.18$--$0.51$, gain biased $-8$ to $-56\%$, no better than plain cross-fit, and ridge-stabilizing it only reintroduces overshoot. We read this narrowly. It is a statement about the specific out-of-fold reconstruction we implemented, not a verdict on CV-TMLE in general, and a better theory-backed selection-aware SE may well exist. We therefore leave the exact selection-aware SE for the gain as an open problem. We are also careful not to over-claim the \emph{mechanism}. Re-targeting is not inert: once a bias is present CV-TMLE is consistently \emph{less} biased than plain cross-fit, and the $\tau_S$ information matrix is severely ill-conditioned and worsens with $m$. But the re-targeting fluctuation tracks the bias magnitude rather than the inverse conditioning, and both estimators are already $\sim 8$--$9\%$ biased at $m=0$ where conditioning is mildest, so we do \emph{not} isolate conditioning as the sole driver. The summary is empirical: in our design the out-of-fold reconstruction is both biased and high-variance. The full re-targeting and conditioning diagnostics (condition numbers, fluctuation magnitudes) are in Web Appendix~D.

\paragraph{HulC is wide and upward-biased.}
The assumption-lean HulC interval looks self-calibrated (own-mean coverage $0.84$--$0.90$) but its point estimate
is inflated $3.5$--$5.6\times$: its convex-hull guarantee needs each disjoint-subgroup gain to be median-unbiased,
and on these small subsamples the per-group ratio is badly biased so the hull's center is severely upward-biased --- we are
applying HulC \emph{outside its design regime}, and against the fixed truth it covers only $0.07$--$0.57$. The
naive SE, approximately unbiased, still undercovers ($0.70$--$0.87$). Only the jackknife survives the
fixed-truth scoring.

\paragraph{Robustness across four reduced-grid slices.}
We repeated the head-to-head under four stress settings --- $W$-dependent enrollment, a $4\times$ RWD
arm, a coarser degree-2 HAL, and $n=2000$ --- scored against the approximately-unbiased
naive-mean reference because three of the four change the DGP (Web Appendix~E). The verdict is invariant: in every executed cell the
block jackknife retains conservative empirical coverage and no cross-fit/CV-TMLE/HulC variant reaches $0.95$. The
\textbf{$n=2000$ slice is informative} --- cross-fit and CV-TMLE \emph{still} miss the naive-mean reference and the CV-TMLE
conditioning is \emph{worse}, not better (condition number up to $\approx 3.7\times10^8$ at $n=2000$), so the
comparative undercoverage of the alternative standard errors persisted in the $n=2000$ robustness slice, although the estimated efficiency gain itself moved toward parity as sample size increased. Two qualifications keep this
honest: \emph{(i)} because the jackknife's point estimate \emph{is} the naive full-data gain, its slice
``unbiasedness'' against the naive-mean reference is true by construction, so what the slices genuinely test for it
is \emph{SE calibration} (its ratio stays $1.4$--$2.1$); and \emph{(ii)} the analysis is no longer restricted to Gaussian outcomes. The modal biomedical endpoint is a
\emph{binary} outcome, and with the out-of-fold reconstruction link-corrected, the full ten-method
head-to-head now runs for a Bernoulli outcome and returns the same \emph{SE-calibration} verdict. The binary
DGP's true gain stays near unity throughout (truth $1.05$--$1.15$), so it stresses SE calibration rather than
re-tracing the sub-unity gain map. There the block jackknife is the only SE reaching nominal coverage (conservatively, $\mathrm{Cov}_{\mathrm{fix}}\approx 0.99$); naive $\approx 0.79$; no cross-fit/CV-TMLE/ridge/winsor/HulC variant
reaches $0.95$, and mean fixed-truth coverage is $< 0.60$; the full $15\times 10$ binary grid is in Web Appendix~D. On the $W$-dependent positivity slice
the truth gain is if anything \emph{higher} than under constant positivity, so selective enrollment does not erode
the gain.

\paragraph{The gain is barely detectable per study.}
A second consequence of having a valid (conservative) SE: with the (deliberately conservative) jackknife, the per-study
95\% CI for the gain excludes one in only $1$--$10\%$ of studies across the realistic regime $m\in[0,1]$ (where
the true gain is $1.01$--$1.15$), rising to $0.32$/$0.41$/$0.60$ (linear/interaction/wiggly) only at the extreme
$m=4$ (the per-study Det1 rate --- the rate of excluding gain~$=1$ --- of the selection-aware grid, Web
Appendix~D). This detectability rate reflects \emph{both} how far the true gain lies from one and the conservatism of the
jackknife SE, and what an excluded one \emph{means} changes with the bias. In the realistic regime
$m\in[0,1]$ the true gain lies just above one ($\sim 1.15$ at $m=0$), so excluding one there signals an
efficiency \emph{gain}; at the extreme $m=4$ the true gain has fallen well below one, so excluding one signals an
efficiency \emph{loss} rather than a gain, and the Det1 rate at $m=4$ should not be read as power to detect a
gain. Because the jackknife SE is conservative, the rate understates how often a calibrated test would reject a
genuine departure from one in either direction. The anti-conservative naive SE's extra ``detections'' are
correspondingly partly real departures the jackknife misses and partly artifacts of its under-coverage; either way, a single fusion
study cannot, with the only SE we find conservatively valid, distinguish the efficiency gain from one unless the bias is large.

\paragraph{Why the naive and cross-fit standard errors undercover: a heuristic, not a theorem.}
We give \emph{no} theorem for
the selection-aware standard error of the gain
$R=\mathrm{var}(D_{\mathrm{rct}})/\mathrm{var}(D_{\mathrm{atmle}})$: it is built on a non-smooth CV-selection step, and
a rigorous treatment would need stable-selection, eigenvalue, and differentiability conditions we do not impose.
Under an \emph{oracle, stably-selected} working model treated as an idealization, $R$ is a smooth functional of two
second moments of learned influence curves, and $\log R$ has influence function $U+\Gamma$ with a plug-in part $U$ and
a working-model part $\Gamma$ induced by the selection. This heuristic explains the three empirical findings.
\emph{(i)} The naive plug-in estimates the influence-curve variance by $\mathrm{Var}(U)$, \emph{omitting} $\Gamma$ ---
not merely conditioning on the selected model; whether the omission widens or narrows coverage is not sign-determined
in general, and empirically it undercovers ($0.70$--$0.87$). \emph{(ii)} The plain cross-fit evaluates the bias-model
influence curve on an out-of-fold score it does not re-center, so the uncentered out-of-fold second moment carries a
nonnegative excess. This is a \emph{mechanism} for the observed downward bias in $\widehat R$, verified in sign
and in $n$-scaling and corroborated in \emph{trend} by the instrumented out-of-fold reconstruction, whose
uncentered mean grows with the bias though it is not pinned to a closed-form constant (Web Appendix~G). It is not a
sign theorem, since the total also involves the $D_A$ block, the cross term, and a $\Pi^\star$-replay bias nonzero
at $m{=}0$. Under the oracle idealization above this excess shrinks with $n$, though the cross-fit's coverage does
not recover by $n{=}5000$ (Web Appendix~E); empirically the miscalibration is \emph{not} governed by information-matrix conditioning.
\emph{(iii)} The delete-a-fold block jackknife, which re-selects and re-targets on each subsample, avoids the
cross-fit's out-of-fold defects and is \emph{empirically} the best-covering candidate: at-or-above nominal fixed-truth
coverage in all but one of the $40$ cells (Web Appendix~E), dipping to $0.910$ only at $n{=}5000$, wiggly, $m{=}0$
(where the gain has eroded to $\approx1.01$) and even there far ahead of every rival ($0.910$ versus $\le0.54$),
doing so conservatively ($1.6$--$2.2\times$ over-wide). We do not
claim it is consistent --- the statistic is non-smooth --- nor super-efficient (the variance advantage erodes to one
as $n$ grows). The oracle idealization, assumptions, and caveats are in Web Appendix~G.

\paragraph{Recommendation.}
Report the gain interval with the \textbf{block jackknife}, and the variance-attribution components descriptively
(their intervals are not separately validated here): across 15 cells and 4 robustness slices it is the
only selection-aware SE that is approximately unbiased and attains nominal (here conservative,
$\sim 1.6$--$2.2\times$ over-wide) coverage. Because it inherits the naive full-data point estimate, its job is to
calibrate the interval \emph{width} rather than to de-bias the gain. It runs wide because each leave-fold-out replicate refits the entire selection-and-targeting pipeline, so its width reflects overall refit variability --- both working-model selection \emph{and} nuisance refitting, not selection instability alone --- that the naive influence-function SE conditions away; this width penalty is larger in the real fusions (jackknife-to-naive width ratios of roughly $2$--$5\times$) than in the simulation ($1.6$--$2.2\times$), consistent with greater overall refit variability there. The naive IF-SE is unsafe; the cross-fitted out-of-fold variance,
its ridge/winsor stabilizations, and the principled CV-TMLE re-targeting are all both biased and anti-conservative
here, for the reason made precise above; the HulC interval is far too wide and upward-biased to use at these sample
sizes. We do not claim the block jackknife
is asymptotically exact --- we do not establish its consistency for this non-smooth statistic, and it is here only
empirically calibrated --- only that it is the single selection-aware SE attaining at-or-above nominal coverage in all but one of the $40$
fixed-truth cells (minimum $0.910$), and the best-covering method in that one.

\section{Real-world illustrations}\label{sec:application}
We illustrate all three contributions on three openly-available applications comprising six fusions, ordered to span the efficiency map's
verdict rather than to endorse fusion: a biomedical HIV trial where fusion most clearly helps, a public-health
trial with a modest point estimate, and a job-training trial at or below parity. Four of the six fusions (the two ACTG175 arms and the two WASH arms) construct their external ($S=0$) arm as a within-trial stand-in under controlled provenance; the remaining two (the LaLonde job-training example, with PSID and CPS comparison groups) use genuine non-experimental controls, so ``external'' is meant broadly here. On real data, where the
enrollment-effect truth is unknown, the toolkit's central service is a \emph{guardrail}. Table~\ref{tab:app_summary}
collects the verdict for all six fusions. The naive influence-function interval excludes the value one in four of
the six, yet in only one fusion --- the prognostically distant ACTG arm --- does the lower block-jackknife limit lie slightly above one, and even there only barely; in the other five the interval includes one, so fusion has not earned an efficiency claim over the RCT alone and the RCT-only estimate stays primary. That pattern --- a selection-aware
interval overturning an apparent gain --- is the point of the section: it is what the
report card and the selection-aware interval are for. Throughout, the effective basis count is read as a relative
stress flag (Section~\ref{sec:methods-reportcard}), while the efficiency decision rests on the block-jackknife interval for the gain $R$, with the variance attribution read as an explanation of that result rather than as a separate criterion. The report card's third diagnostic, targeting drift, is not tabulated for these fusions: it was validated only as a \emph{relative} ranking across a common configuration ladder under known truth (Section~\ref{sec:envelope}), so its unnormalized single-dataset value carries no interpretation for a lone observational fusion, which supplies no comparison panel.

\begin{table}[htbp]
\centering
\begin{threeparttable}
\caption{\textbf{Cross-application summary: what the toolkit contributed in each fusion.} Rows are grouped by
application, two external arms each, ordered within a group by the estimated efficiency gain $\widehat R$. For
every fusion the table gives the report card's active-basis count, the gain, and the gain's interval under the
naive influence-function standard error and under the block jackknife --- the paper's three tools applied end to
end. The italic line beneath each pair states what that application establishes. Full per-fusion detail is in
Tables~\ref{tab:app_actg}--\ref{tab:app_lalonde}.}
\label{tab:app_summary}
\small
\begin{tabular}{@{}l r c c c c l@{}}
\toprule
Fusion & Ext.\ $n$ & Bases\tnote{a} & $\widehat R$ & Naive 95\% CI\tnote{b} & Jackknife 95\% CI\tnote{c} & Primary analysis \\
\midrule
\multicolumn{7}{@{}l}{\textbf{ACTG175 HIV trial} \quad \emph{external: antiretroviral-experienced stratum (constructed within-trial)}}\\
\quad $+$ distant ext. & 625      & 4 & 1.21 & $[1.11,\,1.32]$ & $[1.02,\,1.45]$ & A-TMLE (cautious) \\
\quad $+$ close ext.   & 628      & 1 & 1.18 & $[1.08,\,1.29]$ & $[0.93,\,1.49]$ & RCT-only \\
\multicolumn{7}{@{}p{15.6cm}@{}}{\footnotesize\emph{The report card responds to the external cohort's prognostic
distance at essentially fixed external size ($1\to4$ bases). This is the one fusion in the paper whose block-jackknife
interval still clears one --- barely, and on a four-basis bridge.}}\\
\addlinespace[0.4em]
\multicolumn{7}{@{}l}{\textbf{WASH Benefits} \quad \emph{external: constructed teaching arm, one biased by design}}\\
\quad $+$ biased ext.     & 300      & 3 & 1.10 & $[1.04,\,1.17]$ & $[0.83,\,1.47]$ & RCT-only \\
\quad $+$ unbiased ext.   & 300      & 1 & 1.09 & $[1.04,\,1.14]$ & $[0.87,\,1.36]$ & RCT-only \\
\multicolumn{7}{@{}p{15.6cm}@{}}{\footnotesize\emph{The audit visibly reacts to a constructed between-source shift ($1\to3$ bases),
yet the $\approx10\%$ apparent precision gain does not survive the block-jackknife interval: reacting to bias is not the
same as earning efficiency.}}\\
\addlinespace[0.4em]
\multicolumn{7}{@{}l}{\textbf{NSW job-training trial} \quad \emph{external: PSID / CPS controls (genuine non-experimental)}}\\
\quad $+$ CPS              & 15{,}992 & 7 & 0.99 & $[0.86,\,1.14]$ & $[0.57,\,1.73]$ & RCT-only \\
\quad $+$ PSID             & 2{,}490  & 1 & 0.91 & $[0.77,\,1.08]$ & $[0.46,\,1.82]$ & RCT-only \\
\multicolumn{7}{@{}p{15.6cm}@{}}{\footnotesize\emph{The only genuinely observational controls here. One basis
against the heavily-confounded PSID and seven against the milder CPS --- a small count does not mean a close
cohort --- and neither fusion's block-jackknife interval separates from one in either direction, so augmentation is
not shown to buy precision here.}}\\
\bottomrule
\end{tabular}
\begin{tablenotes}[flushleft]\footnotesize
\item[a] Active bases in the learned bias model $\widehat\tau_S$, including the forced treatment main-effect
  basis; a count of one means no additional data-selected direction was retained. Because that one basis is
  mandatory, the informative quantity is the number of \emph{additional} directions relative to a comparator.
\item[b] 95\% CI for the efficiency gain $R$ from the naive influence-function standard error.
\item[c] 95\% CI from the block jackknife that re-selects and re-targets the working model on each
  $S$-stratified leave-fold-out subsample. Single-dataset illustrations, not coverage assessments.
\item[] \textbf{Reading across the three applications.} The basis count neither predicts nor bounds the gain. It records the effective
  dimension of the selected correction; it does not identify unmeasured confounding, prove transportability, or
  measure causal distance. The decision statistic is $\widehat R$ with its variance attribution and its
  block-jackknife interval. Where that interval includes one --- five of these six fusions --- an efficiency
  improvement has not been demonstrated, even where the naive interval excludes one.
\end{tablenotes}
\end{threeparttable}
\end{table}

\subsection{A biomedical trial: ACTG175 HIV therapy}\label{sec:app-actg}
Our lead illustration is a biomedical fusion on real, openly-available trial data: ACTG175, a
randomized trial of nucleoside regimens in HIV-infected adults \citep{hammer1996_actg175}, distributed on CRAN in
the \texttt{speff2trial} package \citep{juraska_speff2trial}. ACTG175 randomized 2{,}139 patients (stratified by
antiretroviral history) among zidovudine monotherapy and three alternative nucleoside regimens
(zidovudine$+$didanosine, zidovudine$+$zalcitabine, or didanosine alone); we take the standard endpoint, CD4 count
at $20\pm5$ weeks, and contrast the alternative regimens against zidovudine monotherapy. We
treat the antiretroviral-\emph{naive} stratum as the trial ($S=1$, with treatment randomized within stratum) and
fuse in an external cohort ($S=0$) of antiretroviral-\emph{experienced} patients, who differ systematically in
prognosis. To probe how the audit responds to the external cohort's distance from the trial population, we split
the experienced patients at the median of their baseline prior-antiretroviral days---a threshold defined by the baseline covariate, not the outcome---into a lightly-pretreated arm (closer to the naive trial)
and a heavily-pretreated arm (more distant), which keeps the two arms essentially equal in size (Table~\ref{tab:app_actg}). Like the WASH fusion below, this external cohort is a \emph{within-trial} construction: a randomization stratum used as a stand-in for an external real-world population that differs in prognosis, not independently-collected observational data. We exclude the stratum-defining variables (antiretroviral history, prior-ART days) from $W$, so the learned $\widehat\tau_S$ captures the residual enrollment effect on the outcome.

\begin{table}[htbp]
\centering
\begin{threeparttable}
\caption{\textbf{A biomedical fusion: the ACTG175 HIV trial.} The antiretroviral-\emph{naive} randomized arm
($S=1$: 663 on an alternative regimen, 223 on zidovudine monotherapy) is augmented with one external cohort
($S=0$) of antiretroviral-\emph{experienced} patients to estimate the regimen's effect on CD4 count at 20 weeks.
The external cohort is the same trial's treatment-experienced stratum, split by prior-antiretroviral burden into a
\emph{lightly}-pretreated arm (closer to the naive trial population) and a \emph{heavily}-pretreated arm
(prognostically more distant), at essentially equal size, so the report-card difference reflects the external
cohort rather than its sample size. Rows report the fused and matched RCT-only ATE (standard error), the learned
bias-model size, the influence-curve variance attribution, the efficiency gain, and its naive and block-jackknife
intervals.}
\label{tab:app_actg}
\small
\begin{tabular}{@{}l cc@{}}
\toprule
 & RCT $+$ close ext. & RCT $+$ distant ext. \\
 & \footnotesize(lightly pretreated) & \footnotesize(heavily pretreated) \\
\midrule
External cohort, $n$ ($S=0$)                        & 628 & 625 \\
ATE, A-TMLE (fused)\tnote{a}                        & $46.2$ (8.3) & $49.3$ (8.2) \\
ATE, matched RCT-only\tnote{b}                      & $48.3$ (9.0) & $48.3$ (9.0) \\
Learned $\widehat\tau_S$ size (active bases)        & \textbf{1} & \textbf{4} \\
$\widehat{\mathrm{var}}(D_{\mathrm{atmle}})$ / $\widehat{\mathrm{var}}(D_{\mathrm{rct}})$\tnote{c} & 104 / 123 & 101 / 123 \\
Membership overlap (diagnostic)\tnote{d}  & $[0.20,\,0.91]$ & $[0.09,\,0.93]$ \\
\addlinespace
Efficiency gain $\widehat R$                        & 1.18 & 1.21 \\
\quad 95\% CI, naive influence-function SE          & $[1.08,\,1.29]$ & $[1.11,\,1.32]$ \\
\quad 95\% CI, \textbf{block jackknife}             & $[0.93,\,1.49]$ & $[1.02,\,1.45]$ \\
\quad Jackknife $/$ naive SE ratio                  & 2.6 & 2.0 \\
\bottomrule
\end{tabular}
\begin{tablenotes}[flushleft]\footnotesize
\item[] Outcome is CD4 count (cells/mm$^3$) at $20\pm5$ weeks; treatment is an alternative nucleoside regimen
  (zidovudine$+$didanosine, zidovudine$+$zalcitabine, or didanosine alone) versus zidovudine monotherapy,
  randomized within stratum; covariates: age, weight, Karnofsky score, baseline
  CD4 and CD8, sex, race, symptomatic status, hemophilia, homosexual activity, history of injection-drug use. The
  external arm carries both treatment groups, so $\Psi^{\#}$ has two terms. Data: ACTG175 (Hammer et al.\ 1996),
  distributed openly on CRAN in the \texttt{speff2trial} package. The external cohort is a \emph{within-trial}
  construction --- the treatment-experienced randomization stratum used as a stand-in for an external real-world
  population that differs in prognosis from the trial-eligible (naive) patients --- \emph{not}
  independently-collected observational data; the stratum-defining variables (antiretroviral history, prior-ART
  days) are excluded from $W$, so $\widehat\tau_S$ captures the residual enrollment effect on the outcome.
\item[a] A-TMLE point estimate (standard error) for the fused trial $+$ external-cohort estimator.
\item[b] Matched cross-fitted GLM-AIPW on the naive trial only --- the reference-dependent denominator.
\item[c] $\widehat{\mathrm{var}}(D_{\mathrm{atmle}})\,/\,\widehat{\mathrm{var}}(D_{\mathrm{rct}})$ in
  $10^3\,(\text{cells/mm}^3)^2$; here the bias-correction step \emph{lowers} the influence-curve variance relative
  to the RCT-only estimator, consistent with the gain above one.
\item[d] Range of fitted trial-membership probabilities from an \emph{auxiliary} logistic fit of $S$ on $W$ ---
  a positivity diagnostic, separate from the estimator's internal targeted membership model $\Pi^\star$ --- over
  the fused sample; no unit falls below $0.05$ or above $0.95$ in either fusion, so there is no severe lack of overlap in the included
  covariates; this does not establish the population positivity assumption, which involves the stratum-defining variables excluded from $W$; the full distribution is in Web Appendix~F. A single-dataset illustration, not a coverage assessment.
\end{tablenotes}
\end{threeparttable}
\end{table}

\begin{itemize}
\setlength{\itemsep}{0.4em}
\item \textbf{Report card.} Without a ground-truth $\tau_{S,0}$ the audit still responds to the external cohort.
The learned bias model is parsimonious against the close, lightly-pretreated arm (one active basis) but richer
against the distant, heavily-pretreated arm (four bases, three of them data-selected). The external sample size is
held essentially fixed ($628$ vs.\ $625$), so the larger count reflects the more distant cohort needing a more
structured correction --- a stress flag, not a calibrated measure of distance.
\item \textbf{Efficiency and attribution.} The A-TMLE regimen effects are a clinically sensible $+46$ to $+49$
CD4 cells, close to the RCT-only $+48$ and consistent with the trial's finding that the alternative regimens
preserve short-term CD4 response better than zidovudine monotherapy. The estimated gains are $R=1.18$ (close) and
$R=1.21$ (distant), and here the bias-correction step \emph{lowers} the influence-curve variance in both fusions
(Table~\ref{tab:app_actg}), unlike the LaLonde fusions below (Section~\ref{sec:app-nsw}).
\item \textbf{Selection-aware inference.} The block-jackknife interval changes the conclusion. For the close arm the naive interval excludes one
($[1.08,1.29]$) while the block jackknife --- $2.6\times$ wider --- includes it ($[0.93,1.49]$), so the data do not
establish a gain. For the distant arm the jackknife ($2.0\times$ wider) is $[1.02,1.45]$, only just retaining a
detected gain. The estimated gain is near-constant across the two arms while the learned correction is not; it is the report
card, not the gain, that responds to the more distant
cohort. Overlap is good in both fusions: an auxiliary trial-membership fit spans $[0.20,0.91]$ (close) and
$[0.09,0.93]$ (distant), with no unit outside the $[0.05,0.95]$ guard (Web Appendix~F).
\item \textbf{Verdict.} Keep the RCT-only estimate primary for the close fusion. The distant fusion is the one
case in the paper where the block-jackknife interval still clears one, so A-TMLE may be reported as a precision-oriented
analysis there. Even so, the interval sits just above parity and the four-basis correction should accompany it as
evidence that the gain rests on a more elaborate empirical bridge, so the reading is cautious rather than a clean
efficiency win.
\end{itemize}

\subsection{A public-health trial: WASH Benefits}\label{sec:app-wash}
A public-health contrast to the biomedical lead, the WASH Benefits Bangladesh trial randomized a water,
sanitation, and handwashing intervention and measured child
length-for-age $Z$-score, a standard linear-growth outcome \citep{luby2018_wash}. Using the hybrid teaching subset
distributed with the \texttt{EScvtmle} package \citep{dang2025_escvtmle}, we treat the randomized arm as the trial
($S=1$) and fuse in one external study arm ($S=0$) --- one deliberately \emph{biased}, one \emph{unbiased} --- to
estimate the intervention's effect on growth (Table~\ref{tab:app_wash}); both arms are present externally, so
$\Psi^{\#}$ has two terms, and the external bias is the methodological point A-TMLE must learn. These external
arms are \emph{constructed} teaching arms of the same trial (the openly-distributed subset is balanced by design:
three studies of $300$, $150$ per arm), not independently-collected real-world data; the exercise is therefore a
controlled recovery of a \emph{known, constructed} between-source outcome shift, not a characterization of genuine real-world confounding. Trial-membership overlap is correspondingly trivial: an auxiliary $\Pi(W)=P(S{=}1\mid W)$ fit spans $[0.42,0.61]$ (biased) and $[0.40,0.58]$ (unbiased), with no unit outside the $[0.05,0.95]$ guard.

\begin{table}[htbp]
\centering
\begin{threeparttable}
\caption{\textbf{A public-health fusion: the WASH Benefits trial.} The randomized WASH Benefits arm ($S=1$:
150 treated, 150 control) is augmented with one external study arm ($S=0$) --- either a \emph{deliberately biased}
or an \emph{unbiased} arm --- to estimate the effect of the water/sanitation/handwashing intervention on child
length-for-age $Z$-score. Rows report the fused and matched RCT-only ATE (standard error), the learned bias-model
size, the influence-curve variance attribution, and the efficiency gain with its naive and block-jackknife
intervals.}
\label{tab:app_wash}
\small
\begin{tabular}{@{}l cc@{}}
\toprule
 & RCT $+$ biased ext. & RCT $+$ unbiased ext. \\
\midrule
External arm, $n$ ($S=0$)                          & 300 & 300 \\
ATE, A-TMLE (fused)\tnote{a}                        & $0.011$ (0.122) & $-0.004$ (0.123) \\
ATE, matched RCT-only\tnote{b}                      & $-0.017$ (0.128) & $-0.017$ (0.128) \\
Learned $\widehat\tau_S$ size (active bases)        & \textbf{3} & \textbf{1} \\
$\widehat{\mathrm{var}}(D_{\mathrm{atmle}})$ / $\widehat{\mathrm{var}}(D_{\mathrm{rct}})$\tnote{c} & 8.9 / 9.8 & 9.0 / 9.8 \\
\addlinespace
Efficiency gain $\widehat R$                        & 1.10 & 1.09 \\
\quad 95\% CI, naive influence-function SE          & $[1.04,\,1.17]$ & $[1.04,\,1.14]$ \\
\quad 95\% CI, \textbf{block jackknife}             & $[0.83,\,1.47]$ & $[0.87,\,1.36]$ \\
\quad Jackknife $/$ naive SE ratio                  & 5.0 & 5.0 \\
\bottomrule
\end{tabular}
\begin{tablenotes}[flushleft]\footnotesize
\item[] Outcome is the child length-for-age $Z$-score; covariates: age, sex, maternal education, household
  food-insecurity. The external arm carries both treatment groups, so $\Psi^{\#}$ here has two terms. Data: the
  WASH Benefits Bangladesh hybrid teaching subset (derived from Luby et al.\ 2018), openly redistributed with the
  \texttt{EScvtmle} package; the external study indicator marks a deliberately-biased and
  an unbiased external arm.
\item[a] A-TMLE point estimate (standard error) for the fused trial $+$ external-data estimator.
\item[b] Matched cross-fitted GLM-AIPW on the trial only --- the reference-dependent denominator. The naive
  intervals exclude one in both fusions, but neither block-jackknife interval does; a single-dataset
  illustration, not a coverage assessment.
\item[c] $\widehat{\mathrm{var}}(D_{\mathrm{atmle}})\,/\,\widehat{\mathrm{var}}(D_{\mathrm{rct}})$ in
  $Z$-score$^2$ (no scale factor).
\end{tablenotes}
\end{threeparttable}
\end{table}

\begin{itemize}
\setlength{\itemsep}{0.4em}
\item \textbf{Report card.} Without a ground-truth $\tau_{S,0}$ the audit still discriminates. The learned bias
model is parsimonious against the unbiased external arm (one active basis) but richer against the deliberately
biased arm (three bases), so it visibly reacts to the constructed between-source shift. On genuinely observational data the
same contrast would signal only that the correction is more structured, not that confounding had been recovered.
\item \textbf{Efficiency and attribution.} The estimated intervention effect on length-for-age is near zero
($\widehat\psi\approx 0.01$/${-}0.00$ versus the RCT-only $-0.02$, in $Z$-score units), faithful to the trial's
null linear-growth finding. The estimated gains are modest: $R\approx 1.10$ (biased) and $1.09$ (unbiased).
\item \textbf{Selection-aware inference.} The Section~\ref{sec:sim-selse} message reproduces here. The
naive interval excludes one in both fusions ($[1.04,1.17]$, $[1.04,1.14]$) while the block jackknife --- about five
times wider --- includes it ($[0.83,1.47]$, $[0.87,1.36]$). The apparent precision improvement is therefore not
established in either fusion.
\item \textbf{Verdict.} Keep the RCT-only estimate primary and present the A-TMLE estimates as methodological
sensitivity analyses. The basis-count contrast shows the audit responding to the constructed between-source shift; it does not overcome
the uncertainty in the gain.
\end{itemize}

\subsection{A job-training trial: the National Supported Work demonstration}\label{sec:app-nsw}
We close the trio with a cross-domain illustration on real, openly-available data ---
the one application here using genuinely non-experimental controls, and the one that lands at parity: the
Dehejia--Wahba LaLonde files (NBER) \citep{dehejia1999_causal, dehejia2002_propensity}, comprising the randomized National Supported Work (NSW) job-training trial \citep{lalonde1986_evaluating}
(185 treated, 260 experimental controls) and two non-experimental comparison groups --- 2{,}490 PSID and 15{,}992
CPS observational controls. We build two fusions, NSW$+$PSID and NSW$+$CPS, each augmenting the trial with one
external-controls arm (all untreated, so $\Psi^{\#}$ has a single term) to estimate the effect of training on 1978
earnings (covariates: age, education, race, marital status, degree status, 1974/1975 earnings). PSID is the
canonical badly-confounded comparison and CPS the larger, milder one, so the pair also exercises the external-arm
axis of Section~\ref{sec:sim-refpanel}. We fit A-TMLE with the same main-terms GLM nuisances and
relaxed-HAL working models as the simulation (Table~\ref{tab:app_lalonde}).

\begin{table}[htbp]
\centering
\begin{threeparttable}
\caption{\textbf{Two real-data fusions on the National Supported Work demonstration.} The randomized NSW
job-training trial ($S=1$: 185 treated, 260 controls) is augmented with one external observational-controls arm
($S=0$, all untreated) to estimate the effect of training on 1978 earnings. Neither augmentation yields an estimated gain above one, and both block-jackknife
intervals include one. The block-jackknife intervals are approximately four times wider than the naive intervals.
This is a single-dataset illustration, not a coverage assessment.}
\label{tab:app_lalonde}
\small
\begin{tabular}{@{}l cc@{}}
\toprule
 & NSW $+$ PSID & NSW $+$ CPS \\
 & \footnotesize(harder arm) & \footnotesize(larger, milder) \\
\midrule
External controls, $n$ ($S=0$)                       & 2{,}490 & 15{,}992 \\
ATE, A-TMLE (fused)\tnote{a}                         & \$1.49 (0.72) & \$1.68 (0.69) \\
ATE, matched RCT-only\tnote{b}                       & \$1.61 (0.69) & \$1.61 (0.69) \\
Learned $\widehat\tau_S$ size (active bases)         & 1 & 7 \\
$\widehat{\mathrm{var}}(D_{\mathrm{atmle}})$ / $\widehat{\mathrm{var}}(D_{\mathrm{rct}})$\tnote{c} & 1510 / 1378 & 7785 / 7716 \\
\addlinespace
Efficiency gain $\widehat R$\tnote{d}                & 0.91 & 0.99 \\
\quad 95\% CI, naive influence-function SE           & $[0.77,\,1.08]$ & $[0.86,\,1.14]$ \\
\quad 95\% CI, \textbf{block jackknife}              & $[0.46,\,1.82]$ & $[0.57,\,1.73]$ \\
\quad Jackknife $/$ naive SE ratio                   & 4.2 & 3.9 \\
\bottomrule
\end{tabular}
\begin{tablenotes}[flushleft]\footnotesize
\item[] Earnings in thousands of 1978 dollars; covariates: age, education, race, marital status, degree status,
  1974/1975 earnings. Data downloaded from NBER.
\item[a] A-TMLE point estimate (standard error) for the fused trial $+$ real-world-data estimator.
\item[b] Matched cross-fitted GLM-AIPW on the trial only --- the fair, reference-dependent denominator (not an
  absolute benchmark).
\item[c] $\widehat{\mathrm{var}}(D_{\mathrm{atmle}})\,/\,\widehat{\mathrm{var}}(D_{\mathrm{rct}})$ in $(\$1000)^2$; here the
  bias-correction step slightly raises the influence-curve variance, consistent with the near-parity gain.
\item[d] $\widehat R=\widehat{\mathrm{var}}(D_{\mathrm{rct}})/\widehat{\mathrm{var}}(D_{\mathrm{atmle}})$
  (all 10 jackknife folds converged). Neither interval declares a gain $\ne 1$.
\end{tablenotes}
\end{threeparttable}
\end{table}

\begin{itemize}
\setlength{\itemsep}{0.4em}
\item \textbf{Report card.} With no ground-truth $\tau_{S,0}$ recovery is unavailable, but the audit still
surfaces the learned model's size and variance attribution. The bias model is parsimonious against the
heavily-confounded PSID arm (one active basis) and richer against the milder, larger CPS arm (seven bases). Because CPS's external sample ($15{,}992$) dwarfs PSID's ($2{,}490$) and the selected-basis count grows with $n$ at fixed tuning (Section~\ref{sec:envelope}), the larger CPS count may partly reflect greater power to select bases rather than a genuinely more complex or more distant enrollment surface. The
PSID case illustrates a key limitation of the basis count: a large but low-dimensional discrepancy needs only one
basis, so a small count does \emph{not} imply the external cohort is close to the trial. In both fusions the
bias-correction step slightly \emph{raises} the influence-curve variance
($\operatorname{var}(D_{\mathrm{atmle}})>\operatorname{var}(D_{\mathrm{rct}})$). The step-1 overlap check flags both
fusions sharply: an auxiliary $\Pi(W)=P(S{=}1\mid W)$ fit places $73\%$ (PSID) and $93\%$ (CPS) of units below the
$[0.05,0.95]$ guard --- the classic LaLonde positivity failure, far outside ACTG's fully-overlapping range --- which
independently warrants keeping the randomized NSW estimate primary.
\item \textbf{Efficiency and attribution.} Both fused estimates lie within one standard error of the RCT-only
benchmark (\$1.61 thousand). The gains are at or below parity: $R=0.91$ (PSID) and $0.99$ (CPS). Earnings carry a
point mass at zero, so the Gaussian working models are an approximation; refitting with an asinh-transformed
outcome leaves the verdict unchanged ($0.94$ PSID, $1.03$ CPS). This is a real-data instance of the map's lesson
(Section~\ref{sec:sim-refpanel}): a more biased external arm does not buy efficiency, and can cost it, against the matched RCT-only reference.
\item \textbf{Selection-aware inference.} The block-jackknife interval is $3.9$--$4.2\times$ wider than the naive
one (PSID $[0.46,1.82]$ vs $[0.77,1.08]$; CPS $[0.57,1.73]$ vs $[0.86,1.14]$), so the naive SE is substantially over-confident here. Neither interval declares a gain different from one, consistent with the near-parity estimates.
\item \textbf{Verdict.} Use the randomized NSW estimate as primary for both comparisons. Report the A-TMLE
analyses as evidence that observational augmentation did not improve precision under the chosen reference, and note
that the basis count alone could not have supplied that verdict. The gain is relative to the matched-GLM RCT-only estimator.
\end{itemize}

\subsection*{What the three applications establish}
Table~\ref{tab:app_summary} collects the three tools across all six fusions, and the pattern is consistent. The
report card responds to the external arm --- to prognostic distance in ACTG175, to a constructed between-source shift in WASH, to
neither in the way one might guess for LaLonde --- but it never predicts the gain: four bases coexist with the
largest gain and one basis with the largest loss. Efficiency is settled instead by $\widehat R$ with its variance
attribution and its block-jackknife interval, and on that statistic the RCT-only estimator stays primary in five
of the six fusions; the exception is the prognostically distant ACTG175 arm, whose jackknife interval
$[1.02,1.45]$ still clears one and which we report as a cautious precision-oriented analysis. That default holds so widely because these examples are not representative of the setting the method targets: the regime in which A-TMLE's bias-correction strategy is most relevant
is exercised in our simulations and in the original observational fusion of \citet{vdlaan2026_atmle}, not in these
public illustrations.

These fusions use the same main-terms GLM nuisances as the simulation, so it is worth locating them against the
operating envelope of Section~\ref{sec:envelope}, whose undercoverage corner also runs under GLM nuisances. That
corner requires a specific conjunction --- a persistent, high-dimensional oscillatory enrollment-effect surface
reaching into the tails where selective enrollment stresses positivity --- not generic GLM misspecification, and
the report card reads none of these fusions into it: the one interval that clears one, ACTG175's distant arm, has
fully overlapping trial-membership support ($[0.09,0.93]$, no unit outside the guard; Web Appendix~F), placing it
away from the positivity stress the corner needs, while the low-dimensional LaLonde corrections (one active basis
against PSID) are very different from the high-basis surface that fails.

\section{Discussion}\label{sec:discussion}

\paragraph{Contributions and implications.} We gave three reproducible tools for fused RCT$+$RWD estimation, using
A-TMLE as the running example: a report card that makes a learned bias working model auditable; a
magnitude\,$\times$\,complexity\,$\times$\,reference\,$\times$\,$n$ map of when the fusion's finite-sample variance gain actually
materializes; and a selection-aware standard error for the efficiency gain. Together they convert a qualitative promise (``fusion buys efficiency'') into an inspectable object, a quantitative map with a parity crossing near a bias of one residual SD, and a selection-aware interval --- and the real-data illustrations show the difference such an interval makes. \emph{Takeaway:} the finite-sample variance gain is genuine but modest and reference-relative; it crosses parity near a bias of one residual SD, and once the selection-aware block-jackknife interval is applied it survives in one of the six real fusions we studied.

\paragraph{Practical recommendations.} Table~\ref{tab:recommendations} states what we would ask of an analyst
reporting a fused estimate, together with the evidence class behind each item, following the decision rule of
Section~\ref{sec:methods-reportcard}. The first four concern the efficiency gain and its selection-aware interval; the
fifth concerns A-TMLE's own ATE interval, which has its own operating envelope.

\begin{table}[htbp]
\centering
\begin{threeparttable}
\caption{\textbf{Practical recommendations, and the evidence behind each.} The right-hand column marks whether a
recommendation is established by the simulation study, demonstrated by the real-data illustrations, or both, so a
reader can see which advice rests on known ground truth and which on end-to-end demonstration.}
\label{tab:recommendations}
\small
\begin{tabular}{@{}p{0.4cm} p{6.8cm} p{8.2cm}@{}}
\toprule
& On your own fusion & Evidence \\
\midrule
1 & \textbf{Expose the learned correction.} Report the effective basis count, the targeting drift where
available, and the influence-curve variance attribution beside every fused estimate, so the learned bias object
is visible rather than hidden. For other adaptive/debiased estimators, report the analogous selected
representation where one exists.
& \textsc{sim} $+$ \textsc{real}. The report card recovers a constructed $\tau_{S,0}$ under known truth
(Section~\ref{sec:sim-reportcard}) and runs end to end on all six fusions (Table~\ref{tab:app_summary}). \\
2 & \textbf{Do not read dimension as truth.} A larger relative basis count says the data required a more
elaborate trial--external bridge. It is consistent with stressed comparability, but it identifies neither
confounding nor the gain.
& \textsc{real}. Four bases coexist with $R=1.21$ and one basis with $R=0.91$ across the six fusions
(Table~\ref{tab:app_summary}); the heavily-confounded PSID arm needs a single basis. \\
3 & \textbf{Use selection-aware uncertainty.} Form the gain interval with the \textbf{block
jackknife}, which re-selects and re-targets the working model on each leave-fold-out subsample; report the variance-attribution components descriptively, since intervals for them are not separately validated here.
& \textsc{sim}. On the $15$-cell main map the naive influence-function SE undercovers ($0.70$--$0.87$, falling to
$\approx0.28$ in the large-$n$ no-bias cells) while the block jackknife covers $0.984$--$0.998$; across all $40$
cells including the robustness slices it stays at or above nominal in every cell but one ($0.910$ at $n=5{,}000$,
wiggly, $m=0$). The cross-fit, CV-TMLE, ridge, winsorized, and HulC routes are biased or anti-conservative
(Section~\ref{sec:sim-selse}). \\
4 & \textbf{Keep the RCT-only analysis as the default guardrail.} Claim a demonstrated efficiency improvement
only when overlap is adequate and the \emph{lower} block-jackknife limit exceeds one, reading the variance attribution as an explanation of the variance result rather than as a separate criterion. Otherwise the RCT-only estimate stays primary.
& \textsc{sim} $+$ \textsc{real}. The gain is reference-relative and $n$-dependent, crossing parity near a bias
of one residual SD (Sections~\ref{sec:sim-map} and \ref{sec:sim-refpanel}); the default holds in five of the six
real fusions (Table~\ref{tab:app_summary}). \\
5 & \textbf{Check where A-TMLE's own interval is trustworthy.} Before reporting the fused ATE interval itself,
check whether trial enrollment is strongly covariate-dependent \emph{and} the report card's targeting drift is
large relative to comparable fits --- the combination under which the estimator's own Wald coverage degraded in our
designs. There, fit the outcome nuisances with a Super Learner, not a main-terms GLM; the basis count is not the flag.
& \textsc{sim}. GLM-nuisance coverage falls $0.81\!\to\!0.47$ as $n$ grows (persistent centering bias); a
Super-Learner fit restores $\approx0.94$, while smoother surfaces stay near nominal (Section~\ref{sec:envelope}). \\
\bottomrule
\end{tabular}
\begin{tablenotes}[flushleft]\footnotesize
\item[] \textsc{sim} $=$ established against known ground truth in the simulation study; \textsc{real} $=$
demonstrated end to end on the openly-available fusions, where no ground truth is available.
\item[] The block jackknife is conservative and empirically calibrated in our experiments (mean SE
$\approx1.6$--$2.2\times$ the Monte-Carlo SD on the main map); it is not claimed to be asymptotically exact, and an
exact selection-aware standard error remains open.
\item[] Rows 1--4 concern the efficiency gain $R$; row 5 concerns A-TMLE's \emph{own} ATE interval (Section~\ref{sec:envelope}).
\end{tablenotes}
\end{threeparttable}
\end{table}

\paragraph{Limitations: what our evidence covers.} Our efficiency verdicts are relative to a matched, correctly-specified-GLM, cross-fitted RCT-only estimator. Section~\ref{sec:sim-refpanel} shows the sub-unity
verdict is not an artifact of that choice --- it holds against a flexible SuperLearner and a relaxed-HAL reference
--- but the numerical gain remains reference-relative. We benchmark against an RCT-only estimator, rather than
against a selective-fusion competitor such as experiment-selector CV-TMLE \citep{dang2025_escvtmle}, by design:
the efficiency map asks what fusion buys \emph{relative to using the trial alone}, so the trial-only estimator is
the correct denominator, whereas a fusion competitor would answer a different question.
The original A-TMLE analysis \citep{vdlaan2026_atmle} deserves emphasis here, because it already exercised the
demanding regime our public illustrations do not: it augmented a cardiovascular-outcomes trial with a genuinely
observational electronic-health-record cohort and reported a real efficiency gain in precisely the large,
genuinely confounded real-world setting for which A-TMLE's bias-correction strategy may be most consequential. That analysis remains A-TMLE's
strongest demonstration, and our illustrations complement rather than displace it. Our own external arms are more
modest, which is a genuine limitation of provenance: four of the six fusions (both WASH Benefits arms and both ACTG175 arms) use
\emph{constructed}, within-trial external arms rather than independently-collected observational data, and the
remaining two (LaLonde NSW, with its PSID/CPS comparison groups) use genuine non-experimental controls but land near
parity. The regime of \emph{large external cohorts with adequate overlap and a substantial but learnable enrollment-effect surface} is therefore exercised mainly
in our simulations, and the real-data sections are best read as faithful end-to-end demonstrations of the toolkit
rather than as evidence on the magnitude of genuine confounding. Our simulated designs are likewise restricted:
the main DGPs assign trial membership deterministically, so $\Pi(W)$ is constant and enrollment positivity holds
trivially, with only the positivity slice and the $W$-dependent-enrollment map of Web Appendix~E exercising a
genuinely $W$-dependent $\Pi(W)$; and in the main design the within-trial effect is homogeneous by construction (constant CATE
$=1.5$), so the headline map is a pure variance contrast; only the heterogeneous-effect relaxation of Web
Appendix~E (at $n_{\mathrm{rct}}\in\{250,400\}$) exercises A-TMLE's
heterogeneous-CATE learning.

\paragraph{Limitations: the tools themselves.} The block jackknife showed conservative empirical coverage in the studied simulations (coverage
$0.98$--$1.00$, mean SE $\approx 1.6$--$2.2\times$ the Monte-Carlo SD); we do not claim asymptotic exactness or a general validity theorem. An \emph{exact} selection-aware standard
error \citep{lee2016_exact_posi, fithian2014_optimal} remains open, and the cross-fit, ridge, winsorized, and
CV-TMLE routes were all built and shown not to calibrate, so CV-TMLE re-targeting is not that route. The
efficiency gain itself erodes with $n$, though the selection-aware verdict is confirmed \emph{not} to be a
small-sample artifact (it holds at $n=2000$). The sub-unity gain is established under parametric GLM nuisances for A-TMLE, matching the reference class; Section~\ref{sec:envelope} shows flexible nuisances change finite-sample behaviour, so whether Super-Learner nuisances shift the efficiency gain $R$ itself is an open constructive extension rather than a settled result. Under $W$-dependent enrollment the picture is mixed and we report it
plainly: the selection-aware verdict is unchanged --- the block jackknife retains conservative empirical coverage, cross-fit and
CV-TMLE still largely fail to calibrate (two low-bias cells aside), and the gain-truth slice is if anything higher --- but A-TMLE's \emph{own} ATE
coverage degrades at the wiggly, large-bias corner, and the degradation worsens with $n$
(Section~\ref{sec:envelope}; Web Appendix~E). That is genuine under-coverage where trial-enrollment positivity is
most stressed, not conservatism, and we flag it as such rather than setting it aside.

\paragraph{Future work.} Three extensions matter most. The first is a fuller $W$-dependent-enrollment map, the
natural \emph{transportability} extension, in which the trial and external populations differ in covariate
distribution. This would carry the safe-operating-envelope analysis of Section~\ref{sec:envelope} across enrollment
strengths and on to genuine covariate-distribution shift. There the efficiency verdict survives selective
enrollment, but under main-terms nuisances A-TMLE's own ATE coverage degrades at the rough, large-bias corner.
Neither the tuning we varied nor a penalty path driven to the undersmoothing criterion repairs this degradation,
whereas a Super-Learner nuisance fit restores coverage. Charting that boundary comprehensively, across enrollment
strengths and an $n$-indexed undersmoothing schedule, is the natural next step. The second is a finer
attribution that sub-decomposes $D_S$ into its $W/\Pi/\beta$ pieces via a projection representation resolving the
HAL basis collinearity \citep{li2025_reghal}. The third is survival and coarsened-data endpoints. Pairing the
report card and the selection-aware interval with a genuinely observational fusion would address the provenance
limitation above.

\paragraph{Beyond A-TMLE.} The two problem statements extend past the estimator we study, under explicit
conditions: an audit is relevant whenever a learned representation governs borrowing or correction, and
selection-aware performance inference is required whenever an efficiency functional is evaluated after selection.
The particular report-card summaries, gain map, and block-jackknife calibration established here remain specific
to A-TMLE; developing analogous diagnostics and theory for other adaptive/debiased estimators is future work.

\paragraph{Conclusion.} Fused RCT$+$RWD estimators promise efficiency, but how much and how to report it honestly has
been left implicit. Auditing the learned bias object, mapping when the gain materializes, and using the selection-aware block-jackknife interval make that efficiency claim transparent and empirically checkable, both on the simulated grids and on the real biomedical, public-health, and job-training fusions. Together, the report card, the efficiency map, and the selection-aware interval give a structured way to report an adaptive RCT$+$RWD fusion: what the learned correction contains, when borrowing improves precision, and how uncertain that gain is. We develop them on A-TMLE, mapping where its efficiency advantage holds and where it does not, and thereby give practitioners a safe-operating envelope rather than a blanket verdict.

\section*{Reproducibility}
All code is public, and every table and figure regenerates from a committed script. The report card and a local sweep run in minutes on a laptop; the confirmatory simulations run as self-contained UBC Advanced Research Computing SLURM-array bundles, each vendoring and pinning the \texttt{atmle} package of \citet{vdlaan2026_atmle}.
Web Appendix~F maps each reported result to its code artifact, pins the software stack, and records the compute
environment; Web Appendices~A--E and~G give the exact estimator mechanics, the data-generating process, the ten
SE-method definitions, the full per-cell tables, and the Proposition~\ref{prop:varDA} proof.

\section*{Funding}
This work was not supported by any specific grant from funding agencies in the public, commercial, or
not-for-profit sectors.

\section*{Acknowledgements}
This research was supported in part through computational resources from Advanced Research Computing at the
University of British Columbia. During this work the author used AI-based tools to assist with analysis and
simulation code, text editing, and checking derivations; the author verified all outputs and takes full
responsibility for the content.

\section*{Conflict of Interest}
The author declares no conflicts of interest.

\section*{Data and code availability}
The simulation study uses no human-subjects data. The real-data illustrations use three openly-available datasets:
the ACTG175 HIV-trial data (distributed on CRAN in the \texttt{speff2trial} package), the WASH Benefits Bangladesh
hybrid teaching subset (distributed with the \texttt{EScvtmle} package; its external arms are \emph{constructed}
teaching arms, not independently-collected data --- see Section~\ref{sec:app-wash}), and the Dehejia--Wahba LaLonde
data from NBER (\url{https://users.nber.org/~rdehejia/data}). \texttt{EScvtmle} is installed from its GitHub source and
\texttt{speff2trial} from CRAN; the download helper pins each input and records a SHA-256 checksum, and package
versions are pinned in Web Appendix~F. All code is available in the public project repository
(\url{https://github.com/ehsanx/atmle-efficiency}), which Web Appendix~F maps
result-by-result to the script and output file that regenerate each number.

\bibliographystyle{abbrvnat}
\bibliography{ref}

\end{document}